\documentstyle[epsfig,12pt]{article}

\def\Pom{{\bf I\!P}}

\def\lsim{\mathrel{\rlap{\lower4pt\hbox{\hskip1pt$\sim$}}
    \raise1pt\hbox{$<$}}}         

\def\gsim{\mathrel{\rlap{\lower4pt\hbox{\hskip1pt$\sim$}}
    \raise1pt\hbox{$>$}}}         
\newcommand{\dst}{\displaystyle}
\newcommand{\fr}[2]{\frac{{\dst #1}}{{\dst #2}}}

\begin{document}
~\\

\vspace{2cm}

\begin{center}
{\Huge \bf Diffractive production of $S$ and $D$ wave vector mesons
in Deep Inelastic Scattering}\\

\vspace{3cm} {\Large Igor Ivanov}\\

\vspace{1cm}
{\large Novosibirsk University, Russia\\} 
\vspace{5mm}
and\\ 
\vspace{5mm}
{\large Forschungszentrum Juelich, Germany}
\end{center}

\vspace{4cm}
\begin{flushright}

\parbox{8cm}{
Scientific advisors:\\

{\bf Prof.Univ. Dr. J.Speth} (KFA, J\"{u}lich)\\
{\bf Prof. N.N.Nikolaev} (KFA, J\"{u}lich and ITP, Moscow)\\
{\bf Prof. I.F.Ginzburg} (IM, Novosibirsk) }
\end{flushright}

\vspace{2.5cm}
\begin{center}
J\"{u}lich --- 1999
\end{center}

\newpage
~\\

\vspace{2cm}

\begin{center}
{\large\bf Abstract}\\

\end{center}

In this work the production of vector mesons
in diffractive DIS is analyzed, with an emphasis on
the impact of the internal spin structure of a vector meson
upon its virtual photoproduction rate.
For the first time, the full sets of $s$-channel helicity conserving
and violating amplitudes were derived for pure $S$ and $D$
wave vector mesons.
In the course of analytical and numerical investigation,
we found a striking difference between $S$ and $D$
wave vector meson production.
First, in the case of helicity conserving amplitudes,
$D$ wave vector mesons were found to exhibit dramatically different
$Q^2$-behavior due to abnormally large higher twist contributions.
Besides, we found that helicity conserving amplitudes for $S$ and $D$
states are numerically comparable, which makes the physical $\rho$ meson
very sensitive to $D$ wave admixture. Second, the helicity violating
amplitudes in light vector meson production turned out large
both for $S$ and $D$ wave states. In the case of heavy quarkonia,
these amplitudes became suppressed by the non-relativistic motion
for $S$ wave mesons but they were still large for $D$ wave states.

Aiming at producing sort of a manual on the diffractive vector
meson production, we carried out all calculations in a pedagogical 
manner with maximally full presentation of intermediate steps
and accompanied them with detailed qualitative discussion.

\newpage

\tableofcontents

\newpage

\section{Introduction}

In the past 30 years the particle physics theory has proved
numerous times to provide a good, consistent, unified description
of the great variety of nuclear, low and high energy particle
physics experiments.  Being based on the ideas of QFT
applicability, gauge approach to fundamental interactions,
symmetry and naturalness considerations, the Standard Model
managed to explain virtually all phenomena in electromagnetic,
weak and strong interactions, to predict new particles and
effects. Though the questions of fundamental origin lie beyond the
scope of the Standard Model, its precision in description, for
instance, QED phenomena reaches the magnitude of $10^{-10}$.

However, the current situation is not that optimistic in the
domain of strong interactions. The gauge--based formulation ---
the Quantum Chromodynamics (QCD) --- seems to offer reasonably
good description only of the energetic enough processes (more
accurately: only when every vertex involves at least one highly
virtual particle) thanks to the famous asymptotic freedom. The
major difficulty lies in the behavior of the QCD coupling constant:
$\alpha_s(Q^2)$ exhibits infrared growth and becomes
comparable to unity at $Q^2 \sim$ 1 GeV$^2$. The net result is
that the perturbation theory --- the only prolific universal
treatment of various high-energy processes --- fails to give even
qualitative description of low-energy, essentially
non-perturbative phenomena. Additional difficulties arise from the
non-abelian nature of QCD, chiral symmetry breaking,
non-trivial QCD vacuum, instantons etc.

On the other hand, many separate concepts have been developed,
which do not cling to the perturbative QCD (pQCD) and provide
reasonably good description of phenomena in their applicability
regions. The fundamental problem of the theory of strong
interactions is that these heterogeneous approaches do not
match\footnote{ Just a few examples of poor accordance among
various approaches: the quark generated ladder diagrams do not
appear to correspond uniquely to any of experimentally observed
Regge trajectories. Another example is the vague status of
$\alpha_s =$ const BFKL results in true QCD.}. They do not
comprise a unified picture of strong interactions. Given such a
lack of universal, rigorously derived results, one must admit that
the subject of our investigation belongs to the realm of
phenomenology rather than rigorous theory.

\subsection{Diffractive processes and Pomeron}

In the light of these problems, the careful examination of regions
where two or more approaches overlap (or visa versa, where neither
of the concepts exhausts the interaction) are of great interest.
Diffractive Deep Inelastic Scattering (DIS) is exactly one of
these fields.

\begin{figure}[!htb]
   \centering
   \epsfig{file=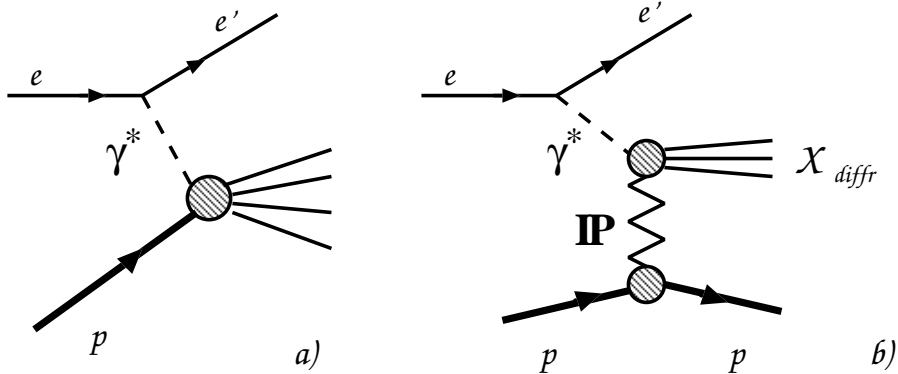,width=120mm}
   \caption{Examples of deep inelastic scattering process:
{\it (a)} hard DIS and {\it (b)} diffractive DIS. In the latter case
$M^2_{diffr} \ll s$ and the process proceeds via pomeron $t$-channel exchange.}
   \label{DIS}
\end{figure}

A typical hard DIS process (Fig.\ref{DIS}a) occurs when a virtual
photon\footnote{We will always imply that the virtual photon is
emitted by an electron, which means the photon is always
space-like: if $q$ is photon momentum, then $Q^2 \equiv -q^2 >0$.}
strikes a proton to produce a hard system $X$ with large invariant
mass\footnote{In hard DIS phenomenology this quantity is usually
labeled as $W^2$. However, for simplicity we will use notation
$s$.} $s$ and large enough multiplicity, final state hadrons being
distributed over whole rapidity range approximately smoothly.

However, as it was noted long ago, sometimes the proton survives,
being only slightly deflected, and a virtual photon turns into a so-called diffractive
system $X_{diffr}$ with invariant mass $M^2_{diffr} \ll s$.
In this process the proton and the diffractive system are naturally
separated by a large rapidity gap and a condition which appears necessary
for the rapidity gap formation is $Q^2 \ll s$, or in terms of Bjorken $x$
\begin{equation}\label{x}
x = {Q^2 \over s} \ll 1\,.
\end{equation}

This is one of the most common cases
of diffractive DIS (DDIS) processes. In fact, the class of diffractive
processes is not confined within DIS; it is much broader.
There are many other reactions which possess the generic features
--- the rapidity gap and smallness of $M^2_{diffr}$ ---
and therefore can be classified as diffractive processes
(for a recent review see \cite{NNN-vocab}).

How can a typical diffractive process occur? Certainly, it must be
kind of a peripheral interaction: if the photon struck directly
one of the valence quarks, the proton would 'explode', providing no
way for the large rapidity gap formation. What remains is the
possibility of the $t$--channel exchange by not-too-energetic
'particle' (Fig.\ref{DIS}b), which would be a natural mechanism of
the experimentally observed  weak proton deflection and small
$M^2_{diffr}$. Further experimental features suggest that this
'particle' should be chargeless and colorless, its interaction
with other particles should be of strong (not EM or weak) nature,
its 'propagation' should be independent of the specific process
($\gamma p$, $\gamma\gamma$, $pp$, $p\bar p$, etc), and it should
be of spin 1 (due to approximately $s$--constant $pp$ cross
section). In the early 60s this 'particle' was dubbed {\bf
Pomeron} (symbol $\Pom$).

Further properties come from combining the Regge picture and BFKL
results with experimental observations (for a detailed review of
Regge theory see \cite{Regge}). They include, first of all, the
asymptotic equality of total $pp$ and $p\bar p$ cross sections
(the Pomeranchuk theorem). Formulated long ago, it was
experimentally verified only recently. Then, the Regge theory
predicts the power-like $s$-dependence of the total $pp$ cross
section $\sigma \propto s^{2(\alpha_\Pom -1)}$, which has also
been experimentally observed, with intercept $\delta_\Pom \equiv
\alpha_\Pom -1 \approx 0.08$. On the other hand, the BFKL equation
\cite{BFKL,BFKLNLO}
succeeded in reproducing such power-like dependence
in QCD, but in a simplified case $\alpha_s = $const. In this
approach the hard pomeron is treated as two reggeized gluons
 --- an ansatz used currently in diffraction phenomenology with great
 success.
However, the predictive power of the BFKL approach for the numerical
value of the pomeron intercept is still limited and
not all issues with sensitivity of the result to  the infrared region
have been understood. For further reading on pomerons,
a topic very intriguing by itself, we refer to \cite{pomerons}.

\subsection{Vector meson production in diffractive DIS}

There are several possible final states $X$ in a typical
diffractive DIS (DDIS) process $\gamma^* p \to X p$:
system $X$ can be a real photon, a $q \bar q$ continuum pair
forming two jets or $q \bar q$ bound state,
for example,  a vector meson.
Let us now focus specifically on exclusive vector meson production
in diffractive DIS.
This reaction has been studied extensively at fixed target
DIS experiments at CERN and FNAL and more recently
by the H1 and ZEUS collaborations at HERA.

In an off-forward scattering
$\gamma^*_{\lambda_\gamma} \to V_{\lambda_V}$
the $s$-channel helicity flip amplitudes can be non-vanishing.
Because of the well known quark
helicity conservation in high energy QCD scattering, such
a helicity flip is possible only due to the internal motion and
spin--angular momentum coupling of quarks in a vector meson.
This issue was accurately analyzed only in very recent papers
\cite{IK,KNZ98}, where it was shown that helicity non-conserving
amplitudes are not negligible, as they had been thought before.
Thus, as such, the helicity flip amplitudes would offer
a great deal of unique information of internal constituent motion
and spin--angular momentum structure of vector mesons,
unaccessible in other experiments. In addition, the vector meson
decays are self-analyzing and the full set of helicity amplitudes
can be measured experimentally. For unpolarized incident leptons,
the angular distribution of decay products is parameterized in terms of
15 spin-density matrix elements, which can be calculated via
5 --- 2 helicity conserving plus 3 helicity violating ---
basic helicity amplitudes \cite{spinmatrix}.

Despite the great deal of theoretical work on vector meson production
in diffractive DIS \cite{IK,KNZ98,early,early1}, the above issue of
sensitivity to the spin--angular momentum coupling has not been
addressed before. Namely, in a typical
vector meson production calculation, a vector meson
has been implicitly taken as $1S$ state and at the same time
an unjustified ansatz was used for $q\bar q \to V$ transition
spinorial structure, namely, of $\bar u' \gamma_\mu u \cdot V_\mu$
type. Being a mere analogy of $q\bar q \gamma$ vertex, this
ansatz in fact corresponds neither to pure $S$ nor to pure $D$
wave state but rather to their certain mixture.
Only in \cite{NNN1s2s} the cases of $1S$ and $2S$ vector mesons
were compared and the necessity of similar calculation for $D$ wave
states was stressed. Such calculations however have been missing in
literature until now.

In addition to purely theoretical needs, there are more issues
that call upon a thorough analysis of the $D$-wave effects. For
instance, different spin properties of the $S$- and $D$-wave
production may resolve the long standing problem of the $D$-wave
vs. $2S$-wave assignment for the $\rho'(1480)$ and $\rho'(1700)$
mesons (as well as the $\omega'$ and $\phi'$ mesons). Furthermore,
the deuteron which is a spin--1 ground state in the $pn$ system is
known to have a substantial $D$ wave admixture, which mostly
derives from the tensor forces induced by pion exchange between
nucleons. Recently, there has been much discussion \cite{Riska} of
the nonperturbative long-range pion exchange between light quarks
and antiquarks in a vector meson, which is a natural source of the
$S$-$D$ mixing in the ground state $\rho$ and $\omega$ mesons.\\

Motivated by these considerations, we undertook a derivation of
a full set of helicity amplitudes for diffractive
electroproduction of pure $S$ and $D$-wave
$q\bar{q}$ systems for all flavors and small to moderate momentum
transfer $\Delta_\bot$ within the diffraction cone.\\

The structure of this thesis is following.
In the next Chapter we develop a QCD--motivated formalism of
pure $S$ and $D$ vector meson treatment. In Chapter 3 a thorough
and detailed derivation of vector meson production amplitudes
is given. In Chapter 4 we continue our analysis in the case
of heavy quarkonia and discuss the most eye-catching differences
between $S$ and $D$ wave production amplitudes.
In Chapter 5 we present some numerical results and finally we
draw conclusions. In the course of calculations,
we aimed at maximally full presentation of intermediate
steps and results. Thus, we hope that the present work might
also be helpful for anyone who intends to start his or her own
calculation in this field. Some of the math--heavy calculations
comprise the Appendix. 

We would like to underline that this work sheds a new light
on the Deep Inelastic Scattering as a means of hadron structure 
investigation.
Indeed, one is accustomed to using DIS as a probe of 
the proton (i.e. target) spin structure.
We demonstrate here that it is a powerful tool for studying 
internal spin structure of a {\it vector meson} as well.
To this point, our work reflects only a few steps in this
promising and still not fully explored domain.\\

All the results given in this thesis were derived by the author.
The only exception was the results shown in 
Fig.\ref{rho0}, which were calculated by
I.Akushevich, Inst. of High Energy Phys., Misnk . 
The majority of the results
derived in this thesis have been published in \cite{IN99} and
presented as talks in \cite{DPG99,DIS99}.

\newpage

\section{Description of a vector meson}

In this section we first introduce the vector meson
light cone wave function (LCWF) and show how it emerges
in diagrammatic calculations. Then, describing $S$ and $D$
wave type vector particles, we give at once expressions for
$S$ and $D$ wave vector meson spinorial structures, which we
then prove by computing the normalization condition for LCWF.
Finally, we also calculate $V \to e^+e^-$ decay constants
to be used afterwards.

\subsection{Bound states in QFT}

While describing particle motion in non-relativistic Quantum Mechanics,
one usually deals with a configuration space particle wave function,
which is a good description because the number of particles is conserved.
So, when one has a system of particles and shows that the wave function
corresponding to their relative motion descreases at large relative
distances at least exponentially, one can speak of a bound state.

In Quantum Field Theory (QFT) this approach needs an update,
since the field function becomes an operator in Fock space.
Besides, since a bound state always implies the presence of interaction,
the projection of a physical bound state onto
the Fock space of {\it free, non--interacting, plane-wave} state vectors
has a rather complicated structure:
\begin{equation}
|V_{phys}\rangle = c_0|q\bar{q}\rangle + c_1 |q\bar{q} g\rangle + c_2 |q\bar{q} gg\rangle
+ c_3 |q\bar{q} q\bar{q}\rangle + ...  \label{fock}
\end{equation}
We emphasize that in this decomposition
quarks and gluons are assumed free, i.e. {\bf on mass shell}.
Coefficients $c_i$ can be called
'wave functions' of the given projection of a physical vector meson,
with $|c_i|^2$ being the probability of finding a vector meson
in a given state.

\subsection{LCWF and vertex factor}\label{sectionLCWF}

Let us now outline how a wave function of a bound state appears in
the diagrammatic language.

In the non-relativistic quantum mechanics, the two-particle
bound state problem can be immediately reformulated
as a problem for one particle of reduced mass $\mu$, moving
in the external potential. This reformulation allows one to
split the wave function into two factors: the wave function
of the motion of the composite particle as whole and the wave function
corresponding to the internal motion of constituents.
The former part factors out trivially, while the latter
wave function obeys the following Schodinger equation
\begin{equation}
\left[{\hat p^2 \over 2\mu} + V(r)\right]\psi(r) = E \psi(r)\,.\label{LCWF1}
\end{equation}
Since the wave function $\psi(r)$ and the interaction operator $V(r)$
exhibit good infinity behavior, one can rewrite this equation
in the momentum representation 
\begin{eqnarray}
&&{p^2 \over 2\mu}\psi(p) + {1 \over (2\pi)^3} \int d^3k V(k)\psi(p-k)
= E\psi(p)\,;\nonumber\\
&&\left({p^2 \over m} - E\right)\psi(p) =
- {1 \over (2\pi)^3} \int d^3k V(k)\psi(p-k)\,.\label{LCWF2}
\end{eqnarray}
In this notation, this equation can be viewed as a homogeneous 
non-relativistic Bethe-Salpeter equation for the wave function 
$\psi(p)$ that describes the relative motion of constituents inside
a composite particle. 

\begin{figure}[!htb]
   \centering
   \epsfig{file=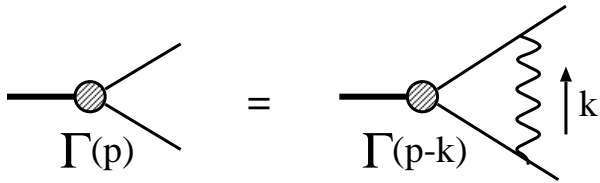,width=80mm}
   \caption{The diagrammatic interpretation of the integral equation
for vertex function $\Gamma(p)$ ($p$ is the relative constituents momentum). }
   \label{LCWFfig}
\end{figure}

Let us now introduce
\begin{equation}
\Gamma(p) \equiv \left({p^2 \over m} - E\right)\psi(p) \label{LCWF3}
\end{equation}
Then Eq.(\ref{LCWF2}) can be rewritten as
\begin{equation}
\Gamma(p) = - {1 \over (2\pi)^3} \int d^3k V(k)
\fr{1}{\fr{(p-k)^2}{m} - E} \Gamma(p-k)\label{LCWF4}
\end{equation}
This equation has an absolutely straightforward diagrammatic interpretation
(Fig.\ref{LCWFfig}). One sees that $\Gamma(p)$ stands for
bound-state $\to$ constituents transition  vertex,
with $p$ being the relative momentum of the constituents,
factor $ 1/[{(p-k)^2 \over m} - E]$ describes propagation of $q\bar q$
pair and $V(k)$ stands for the interaction between constituents.
Of course, the kinetic energy $p^2/2\mu \not = E$, the total energy,
which is in fact negative, so no pole arises in the propagator.

In the relativistic case, i.e. in QFT, it is not clear 
{\it a priori} whether the whole picture that involves 
somehow defined wave function and representation of
the vector meson as free non-interacting constituents
would work at all. So, in our approach we will be aiming at 
introducing an appropriately defined wave function
and demonstrating that hard processes involving vector mesons
can be expressed in terms of expectation
values of $q \bar q$ amplitudes between wave functions, 
i.e. we intend to treat a hard process in a 
{\it probabilistic, quantum mechanics-like manner}.

In the following we will show that this program succeeds.
Namely, we will introduce the {\it radial} wave function 
of the $q \bar q$ state of a vector meson as
\begin{equation}
\psi(q)\equiv  {\Gamma(q) \over M^2 - m_V^2}\,
\label{LCWF6}
\end{equation}
(the angular dependence of the wave function will be treated 
separately, see Sect.\ref{sectspin})
Here $\Gamma(q)$ is the vertex factor,
$M^2$ is the eigenvalue of the relativistic kinetic operator
of the on mass shell $q \bar q$ state and $m_V^2$ is eigenvalue
of the total relativisitic Hamiltonian, which is of course
equal to the mass of the vector meson squared.
Then, during an {\it accurate and honest} analysis of a hard process Feynman
diagrams, we will always make sure that wave function (\ref{LCWF6}) automatically appears
in calculations and {\it the rest} looks the same as if both fermions were 
on mass shell. If we see that fermion virtualities modify the results,
or if different Fock states mix during hard interactions of the vector meson,  
it would signal the invalidity of the free particle parton language 
and consequenly the breakdown of the whole approach. This restriction 
must always be taken into account when obtaining and interpreting the 
parton model-based results.

\subsection{Light cone formalism}\label{sectLC}

The term "light cone approach" to high--energy process
calculations can have different meanings. Some prefer
to re-formulate the whole QFT within light cone dynamics,
introduce light cone quantization and derive light cone Feynman
rules (on Light Cone Field Theory see \cite{LCQFT0,LCQFT}).
 However, one should keep in mind that even
within the usual QFT the light cone formalism can be freely used
as a means to greatly simplify intermediate calculations.
This is exactly the way we will use it.

It was noted long ago \cite{sudak}
that the calculation of a high energy collision
is simplied if one decomposes all momenta in terms of light cone
$n_+^\mu, n_-^\mu$ and transversal components, which we will 
always mark with the vector sign over a letter
(so called Sudakov's decomposition):
\begin{eqnarray}
&&n_+^\mu = {1 \over \sqrt{2}}(1, \vec 0, 1)\,;
\quad n_-^\mu = {1 \over \sqrt{2}}(1, \vec 0, -1)\,;
\quad (n_+n_-)=1\,,\ (n_+n_+) = (n_-n_-) = 0\nonumber\\
&&p^\mu = p_+n_+^\mu + p_-n_-^\mu + \vec p^\mu\,;
\quad p^2 = 2p_+p_- - \vec p^2\,. \label{LC1}
\end{eqnarray}
Indeed, imagine two high energy particles colliding with momenta
$p^\mu$ and $q^\mu$ respectively and equal masses $m$.
Then one can choose such a frame of reference that
in the Sudakov's decomposition
\begin{eqnarray}
&&p^\mu = p_+n_+^\mu + p_-n_-^\mu + \vec p^\mu\,,\nonumber\\
&&q^\mu = q_+n_+^\mu + q_-n_-^\mu + \vec q^\mu\nonumber\label{LC2}
\end{eqnarray}
quantities $p_-$ and $q_+$ are large $p_-,\, q_+ \gg m$ while
$p_+ = (\vec p^2 + m^2)/(2p_-)$ and $q_- = (\vec q^2 + m^2)/(2q_+)$
are small. The total energy squared of these two particles
will be defined as
\begin{equation}
s \equiv 2q_+p_-\,. \label{LC3}
\end{equation}
Note that our definition of $s$ is somewhat different from
the more familiar one $(p+q)^2$ by terms 
$\propto m^2, \vec p^2$. However, it is not of any imporatnce
for us, since in the course of all calculations
we will keep track only of the highest power $s$ contributions,
i.e. everything will be calculated in the leading power $s$ approximation.

\begin{figure}[!htb]
   \centering
   \epsfig{file=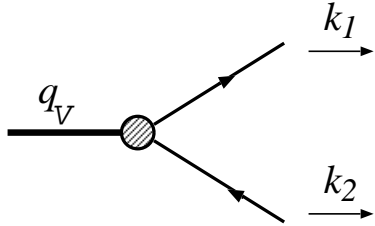,width=50mm}
   \caption{Kinematics of $V \to q\bar q$ vertex on the light cone.
Vector meson momentum $q_V$ is taken incoming, constituents momenta are
outgoing.}
   \label{LCWFkinem}
\end{figure}

Let us now go further and examine the kinematics of
a typical $q \bar q V$ vertex (Fig.\ref{LCWFkinem}).
The Sudakov's decomposition of all momenta reads:
\begin{eqnarray}
&&q_V^\mu = q_{V+}n_+^\mu + q_{V-}n_-^\mu\,;\nonumber\\
&&k^\mu_1 = k_{1+}n_+^\mu + k_{1-}n_-^\mu + \vec k^\mu =
zq_{V+}n_+^\mu + y q_{V-}n_-^\mu + \vec k^\mu\,;\nonumber\\
&&k^\mu_2 =  k_{2+}n_+^\mu + k_{2-}n_-^\mu - \vec k^\mu =
(1-z)q_{V+}n_+^\mu + (1-y)q_{V-}n_-^\mu - \vec k^\mu\,\label{LC6}
\end{eqnarray}
so that
\begin{equation}
q_V^2 = 2q_{V+}q_{V-} = m_V^2\,;\quad
k_1^\mu + k_2^\mu = q_V^\mu\,;\quad k_i^2 \not = m^2\,,
  \label{LC6a}
\end{equation}
i.e. quarks can be off mass shell.
Now let us introduce momenta $k_i^*$, which would correspond to
on mass shell fermions. The only component in $k_i$ subject to modification 
is $k_{i-}$ component, or absolutely equivalently, the energy.
Large $k_{i+}$ components are insensitive to (reasonable) quark virtuality
variations. So, to obtain the on mass shell momenta, one has to replace
\begin{equation}
k_{i-} = {k_i^2 + \vec k^2 \over 2k_{i+}} \to
k^*_{i-} = {m^2 + \vec k^2 \over 2k_{i+}}\,.
  \label{LC6b}
\end{equation}
Then the 4-vector 
\begin{equation}
  \label{LC6c}
  q^\mu = k^{*\mu}_1 + k^{*\mu}_2 
\end{equation}
squared is equal to
\begin{equation}
  \label{LC6d}
  M^2 = q^2 = 2q_+q_- = 2q_{V+}\left(k^*_{1-} + k^*_{2-}\right) = 
{\vec k^2 + m^2 \over z(1-z)}\,.
\end{equation}
And again we emphasize that the {\it Feynman invariant mass} 
(i.e. the total 4-momentum squared) of the virtual 
quark-antiquark pair is $m_V^2$. The quantity $M^2$ is the invariant
mass of the free, non-interacting $q\bar q$ state (see \ref{LCWF6}).
However, it is precisely $M$, not $m_V$ that will govern
the hard interaction of $q\bar q$ pair with gluons.

Finally, it is useful to introduce the relative momentum of 
free $q\bar q$ system: 
\begin{equation}
2p_\mu = (k^*_1 - k^*_2)_\mu\;.\quad
\end{equation}
Then, trivial algebra leads to
\begin{equation}
 M^2 = 4m^2 + 4{\bf p}^2\;;\quad p^2 = - {\bf p}^2\,;
\quad (pq) =0\,,\label{LC10}
\end{equation}
where ${\bf p}$ is the 3--dimensional relative momentum in the
$q\bar q$ pair rest frame of reference. 
Its components are
\begin{equation}
{\bf p} = (\vec p, p_z)\,;\quad \vec p = \vec k\,; \quad 
p_z = {1 \over 2} (2z-1)M\,.\label{LC11}
\end{equation}

\subsection{Spin structure of a vector particle}\label{sectspin}

Let us start with a well known example of a deuteron, which
is a non-relativistic analogy of a vector meson: they are both
vector particles built up of two fermions.
To have the correct $P$-parity, proton and neutron must sit
in the spin--triplet state, thus leaving us with two possible
values of their angular momenta: $L =0$ and 2.

In the conventional non-relativistic language one describes
the spin-angular coupling by the Clebsh-Gordan technique.
The non-relativistic Feynman diagram calculations
can be best performed in an alternative approach.
Here a deuteron, being a vector particle, is described by
a 3 dimensional polarization vector ${\bf V}$.
So, while calculating high energy processes
involving $d \to pn$ transitions,
one can use the following spin structure of
deuteron-nucleon-nucleon vertex:
\begin{equation}
\phi_n^+\,{\bf \Gamma}\, \phi_p \cdot {\bf V} \label{deuteron1}
\end{equation}
Since both nucleons can be treated on mass shell,
only two terms enter $\Gamma_i$, which can be written as:
\begin{equation}
\phi^+_n\,\left[u(p)\sigma^i +
w(p)(3p^ip^j - \delta^{ij} p^2) \sigma^j\right]
\, \phi_p \cdot V^i \label{deuteron2}
\end{equation}
Here $\sigma^i$ are Pauli matrices and ${\bf p}$ is 
the relative proton--neutron momentum.
One immediately recognizes here spin structures
corresponding to $pn$ pair sitting in
$S$ and $D$ waves respectively.
In particular, squaring the above expression
gives
\begin{eqnarray}
&&({\bf V}{\bf V}^*) \quad \mbox{for $|S|^2$} \nonumber\\
&&3({\bf p}{\bf V})({\bf p}{\bf V}^*) -({\bf V}{\bf V}^*){\bf p}^2
\quad \mbox{for $SD$ interference}\nonumber\\
&&3{\bf p^2}({\bf p}{\bf V})({\bf p}{\bf V}^*) +({\bf V}{\bf V}^*){\bf p}^4
\quad \mbox{for $|D|^2$} \label{deuteron3}
\end{eqnarray}

Now, let us go relativistic and turn to vector mesons.
The polarization state of a vector particle is described by
a four-vector $V_\mu$. Therefore, a general form of $q\bar q V$ vertex
has the form
$$
\bar u' \Gamma_\mu u \cdot V_\mu \cdot \Gamma(p)\,,
$$
where $\Gamma(p)$ is the familiar vertex factor.
Up to now, it has been customary in literature 
to choose the simplest form
of the spinorial structure $\Gamma_\mu$:
\begin{equation}
\bar u' \gamma_\mu u \cdot V_\mu \cdot \Gamma(p)\,.\label{naive}
\end{equation}
However, one must admit that (\ref{naive}) is simply an analogy of
$q \bar q \gamma$ vertex and does not reflect the true internal structure
of a vector meson.
It is known \cite{SD} that the correct spinorial structure
corresponding to pure the $S$ wave $q\bar q$ state reads
\begin{equation}
  S_\mu = \gamma_\mu - { 2 p_\mu \over M + 2m} =
\left( g_{\mu\nu} - {2p_\mu p_\nu \over m (M+2m)}\right) \gamma_\nu
\equiv {\cal S}_{\mu\nu}\gamma_\nu\,.\label{Swave}
\end{equation}
It is implied here that spinorial structures are inserted between
{\it on mass shell spinors} in accordance with 
our principal guideline (see discussion in Sect.\ref{sectionLCWF}).

Once $S$ wave spinorial structure
is established, the expression for $D$ wave can be obtained by
contracting $S$ wave with the symmetric traceless tensor of rank two
$3p_ip_j - \delta_{ij}{\bf p}^2$, rewritten in the Lorenz notation.
To do so, one should replace
$$ 
p_i \to p_\mu\,;\quad
\delta_{ij} \to - g_{\mu\nu} + {q_\mu q_\nu \over M^2} 
$$
(in the $q\bar q$ pair rest frame of reference $q_\mu = (M,\,0,\,0,\,0)$).
However, since $q_\mu$ inserted between on mass shell spinors gives zero
due to the Ward identity, one obtains the required tensor
in the form $3p_\mu p_\nu + g_{\mu\nu}{\bf p}^2$.
Its contraction with ${\cal S}_\mu$ yields
\begin{equation}
  D_\mu = (3p_\mu p_\nu + g_{\mu\nu}{\bf p}^2) \cdot {\cal S}_{\nu\rho}\gamma_\rho
= {\bf p}^2\gamma_\mu + (M+m)p_\mu =
\left( {\bf p}^2 g_{\mu\nu} + {M+m \over m}p_\mu p_\nu\right) \gamma_\nu
\equiv {\cal D}_{\mu\nu}\gamma_\nu\,.\label{Dwave}
\end{equation}
We will prove below that structures (\ref{Swave}), (\ref{Dwave})
after being squared indeed perfectly reproduce
(\ref{deuteron3}), i.e. they indeed correspond to pure $S$ and $D$ waves.\\

The quantities ${\cal S}_{\mu\nu}$ and ${\cal D}_{\mu\nu}$
used in (\ref{Swave}), (\ref{Dwave})  have the meaning of
$S/D$ wave projectors, which will be used in all subsequent calculations.
 Namely, all calculations will be at first
performed for the naive $q\bar q V$ vertex (\ref{naive}) and
then we will apply the projector technique to obtain
expressions for $S$ and $D$ wave states.

\subsection{Vector meson LCWF normalization}\label{sectnorm}

Before tackling the diffractive vector meson production process,
we first should have a prescription of normalization of the
vector meson wave function.

\subsubsection{Naive $q\bar q V$ vertex}

\begin{figure}[!htb]
   \centering
   \epsfig{file=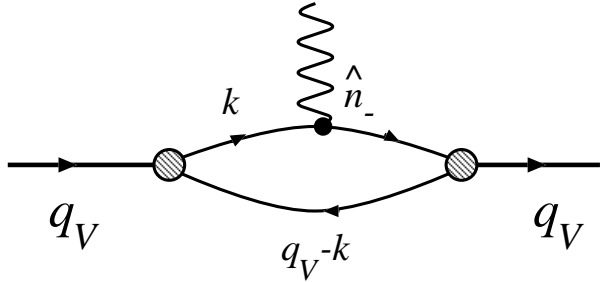,width=80mm}
   \caption{Diagram used for normalizing the vector meson LCWF.
The amplitude of this diagrams is set equal to $2q_+i$.}
   \label{normalization}
\end{figure}

A natural way to normalize the wave function of a composite system is
to put the amplitude given by this diagram in Fig.\ref{normalization}
equal to $2 q_+ i$. Here extra leg carries zero momentum but couples
to the fermion line as $\gamma_\mu n_-^\mu$.
Note that for a charged composite particle
(a deuteron) this is precisely setting the electric formfactor
equal to unity in the soft photon limit. \\

As described above, we first treat $q \bar q V$ vertex as
 $\bar{u}' \gamma_\mu u \cdot \Gamma(p)$. In this case the general expression
for this amplitude is
\begin{eqnarray}
A = { (-1) \over (2\pi)^4} N_c \cdot
\int d^4 k { Sp\{ i \hat V_1 \Gamma \cdot i(\hat k - \hat q_V + m) \cdot
i \hat V^*_2 \Gamma^* \cdot i(\hat k + m) \cdot i \hat n_- \cdot
i(\hat k + m)\} \over [k^2 -m^2 +i \epsilon] \cdot [k^2 -m^2 +i \epsilon]
\cdot [(k-q_V)^2 -m^2 +i \epsilon]} \nonumber\\
= {N_c \over (2\pi)^4}\cdot \int d^4 k { |\Gamma|^2 
Sp\{...\} \over [k^2 -m^2 +i \epsilon]^2 [(k-q_V)^2 -m^2 +i \epsilon]}\,,
\end{eqnarray}
where $N_c=3$ is a trivial color factor originating from the quark loop.
We deliberately recognized $V_1$ and $V_2$ as distinct entities
just to make sure later that such a loop is indeed diagonal 
in polarization states. 

The first step is to rewrite this expression in terms of Sudakov's variables.
As usual, one implements decomposition
$$
k^\mu = zq_{V+}n_+^\mu + y q_{V-}n_-^\mu + \vec k^\mu\,;\quad 
q_V^2 = 2q_{V+}q_{V-} = m_V^2
$$
and transforms
$$
d^4k = {1 \over 2}m_V^2 d^2\vec k dy dz\,.
$$
Now we note that vertex functions $\Gamma$ do not depend on $y$
(and neither does the trace, as will be shown later),
so we can immediately perform the integrations over $y$
by means of Cauchy theorem. Indeed, since the integral
\begin{equation}
\int_{-\infty}^{\infty} dy
{1 \over [yzm_V^2 - (\vec k^2 +m^2) + i \epsilon]^2}
{1 \over [(1-y)(1-z)m_V^2 - (\vec k^2 +m^2) + i \epsilon]}\,,
\label{int}
\end{equation}
is convergent and has good infinity behavior,
one can close the integration contour in the most convenient way.
To do so, one should analyze the position of all poles on the complex
$y$ plane:
$$
y_{1,2} = {\vec k^2 + m^2 \over zm_V^2} - {i\epsilon \over zm_V^2};\quad
y_3 = 1 - {\vec k^2 + m^2 \over (1-z)m_V^2} + {i\epsilon \over (1-z)m_V^2}\,.
$$
One sees that if $z<0$ or $z>1$, {\it all} poles lie on the same
side of the real axis in the complex $y$ plane, which leads to
zero contribution. The contribution that survives comes from
region $0<z<1$, which has in fact a simple physical meaning:
all constituents must move in the same direction.
In this region, we close the integration contour through the
upper half-plane and take a residue at $y = y_3$.
Physically, it corresponds to putting the antiquark on mass shell.
After this procedure, one gets for (\ref{int}):
\begin{eqnarray}
&&- {2\pi i \over (1-z)m_V^2}
{1 \over [yzm_V^2 - (\vec k^2 +m^2) + i \epsilon]^2}\bigg|_{y=y_3}
= -{2\pi i \over (1-z)m_V^2}
{(1-z)^2 \over [\vec k^2 + m^2 - z(1-z)m_V^2]^2}\nonumber\\
&&= - {2\pi i \over m_V^2}{1 \over z^2(1-z)}
{1 \over [M^2 - m_V^2]^2}\,.\nonumber
\end{eqnarray}
One immediately recognizes here the same two-particle propagator
as in (\ref{LCWF6}). Therefore, the equation for the amplitude reads
\begin{equation}
A = i {N_c \over (2\pi)^3} \cdot \int d^2 \vec k {dz \over z^2 (1-z)} \cdot |\psi|^2
\cdot \left(-{1 \over 2} Sp\{...\}\right)\,.
\label{a1}
\end{equation}

Now we turn to the trace calculation. To do this, we will first
show that thanks to the presence of $\hat n_-$
vertex, {\it both constituents} indeed can be treated in the numerator 
algebra as if they were on mass shell.
Indeed, applying the Sudakov decomposition to the $\gamma$ matrix, one has
\begin{eqnarray*}
&&\gamma^\mu = \gamma_+n_+^\mu + \gamma_-n_-^\mu + \vec \gamma^\mu\,;\\
&&\gamma_+ = \hat n_- = {1 \over \sqrt{2}}(\gamma_0 + \gamma_3)\,,\quad
\gamma_- = \hat n_+ = {1 \over \sqrt{2}}(\gamma_0 - \gamma_3)\,.
\end{eqnarray*}
Now let's decompose the propagator numerator of
the constituent, to which this $\hat n_-$ leg couples:
\begin{equation}
\hat k + m = k_+\gamma_- + k_-\gamma_+ - \vec k\vec \gamma +m
\label{propag}
\end{equation}
and rewrite it using notation of (\ref{LC6}),(\ref{LC6b}) as
\begin{equation}
\hat k + m = k_+\gamma_- + k^*_-\gamma_+ - \vec k\vec \gamma +m
+ (k_- - k_-^*)\gamma_+ =
\hat k^* + m + {k^2 - m^2 \over 2k_+}\gamma_+\,.
\label{propag1}
\end{equation}
In other words, we expressed the virtual quark propagator as the sum of
{\it on shell} quark propagator and an additional "instantaneous
interaction" term.
However, since $\hat n_-$ is inserted between two $(\hat k +m)$ factors,
this item does not work due to identity $\gamma_+\gamma_+ = 0$.
The net result is that
{\it wherever $\hat n_-$ appears, both constituents can be treated
on mass shell in the trace calculation}.
This observation is very important for us, since, as we remember,
it validates the whole approach.

Having established this property, we can now easily calculate
the trace. In our case, the easiest way is to do it covariantly,
without involving further the Sudakov technique.
Since quarks in numerator can be treated on mass shell, we first note that
$$
(\hat k + m) \hat n_- (\hat k + m) = 2 (k^* n_-) (\hat k^* + m)
$$
so that
$$
-{1 \over 2} Sp\{...\} = - z q_+ Sp\{\hat V_1 (\hat k^* - \hat q + m)
\hat V_2^* (\hat k^* +m)\} =
-2 z q_+ \left[ M^2 (V_1 V_2^*) + 4 (V_1 p)(V^*_2 p) \right]\,,
$$
where $p$ is the relative quark-antiquark momentum [see (\ref{LC10})].
Note that in the antiquark propagator we replaced $\hat k - \hat q_V 
\to \hat k^* - \hat q$, since the antiquark is now put on mass shell.
Besides, we explicitly used here gauge condition $(qV)=0$, which
means that polarization vectors must be written 
for {\it on mass shell $q\bar q$ pair}, not the vector meson,
--- another important consequence of our approach. 
Substituting this into (\ref{a1}), one gets
\begin{equation}
1 = {N_c \over (2\pi)^3} \int d^2 \vec k {dz \over z(1-z)} |\psi|^2
\left[-M^2 (V_1 V_2^*) - 4 (V_1 p)(V^*_2 p)\right]\,.
\label{a2}
\end{equation}
A prominent feature of this equation is the orthogonality of
$V_L$ and $V_T$ polarization states --- the necessary condition for any
normalization prescription.

The next step is to realize that the integral can be cast in the form of
$d^3 {\bf p}$ integration by means of
$$
{dz \over z(1-z)} d^2\vec k= {4 \over M} dp_z d^2\vec p
= {4 \over M} d^3 {\bf p}\,.
$$
Tus, the final expression for normalization condition is
\begin{equation}
1 = {N_c \over (2\pi)^3} \int d^3 {\bf p}
 {4 \over M}|\psi|^2 [-M^2 (V_1 V_2^*) - 4 (V_1 p)(V^*_2 p)]\,.
\label{a3}
\end{equation}
We see that the expression being integrated is explicitly spherically
non-symmetric, which is a manifestation of a certain $D$ wave admixture.
Thus we now apply projector technique to obtain results
for $S$ and $D$ wave states.

\subsubsection{Normalization for $S$ wave vector meson}

The correct expressions for pure $S$/$D$ type vertices
can be readily obtained with the aid of projector technique.
Namely, to obtain an expression for $S$ wave, replace
\begin{equation}
V_\mu \to V_\nu {\cal S}_{\nu\mu}\,.
\end{equation}
Such a replacement for $V_1$ leads to
$$
-M^2(V_1 V_2^*) - 4 (V_1 p)(V^*_2 p) \Rightarrow
 - M^2(V_1 V_2^*) - {4 M \over M+2m} (V_1 p)(V^*_2 p)
$$
Then, one applies the same replacement to $V^*_2$ to obtain
$$
 - M^2(V_1 V_2^*) - {4 M \over M+2m} (V_1 p)(V^*_2 p)
\Rightarrow - M^2(V_1 V_2^*) \,.
$$
Therefore, the answer for $S$ wave states reads
\begin{equation}
\framebox(200,40){$\dst 1 = \fr{N_c}{(2\pi)^3} \int d^3 {\bf p}\ 
4M |\psi^S({\bf p}^2)|^2\,$}
\label{a6}
\end{equation}
which is manifestly spherically symmetric.

\subsubsection{Normalization for $D$ wave vector meson}

Results for $D$ wave states are derived in the same way. The
replacements $V_\mu \to V_\nu {\cal D}_{\nu\mu}$ lead to
\begin{equation}
-M^2{\bf p}^4(V_1 V_2^*) + 3M^2{\bf p}^2 (V_1 p)(V^*_2 p)\,.
\label{a8}
\end{equation}
After angular averaging
$$
\langle p_i p_j\rangle \to {1\over 3} {\bf p}^2 \delta_{ij},
$$
one gets the normalization formula for $D$ wave state:
\begin{equation}
\framebox(200,40){$ \dst 1 = \fr{N_c}{(2\pi)^3} \int d^3 {\bf p}\ 8M {\bf p}^4
|\psi^D({\bf p}^2)|^2\,$}\label{a9}
\end{equation}

Several remarks are in order. First, $S$ wave $\to$ $D$ wave transitions 
are forbidden. Indeed, such an amplitude
will be proportional to
\begin{equation}
-M^2[{\bf p}^2(V_1 V_2^*) + 3(V_1 p)(V^*_2 p)]\,.\label{a10}
\end{equation}
which vanishes after angular integration.
Then, we emphasize that the structure of results
(\ref{a6}), (\ref{a8}), (\ref{a10}) is absolutely identical to Eq.(\ref{deuteron3}).
This fact can be viewed as the {\it proof} that spinorial structures
(\ref{Swave}), (\ref{Dwave}) indeed correspond to pure $S$ and $D$ wave states.

\subsection{Decay constant}

\begin{figure}[!htb]
   \centering
   \epsfig{file=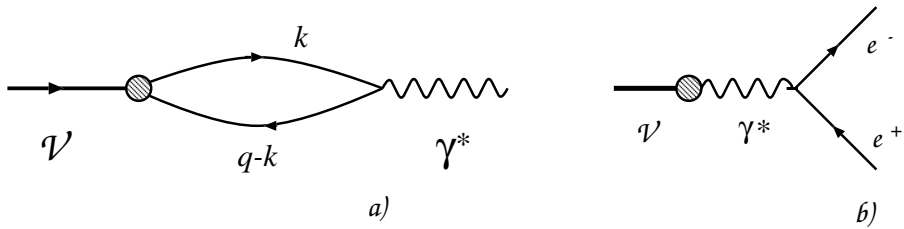,width=120mm}
   \caption{Normalizing LCWF to $\Gamma(V\to e^+e^-)$ decay width:
 (a) the diagram for $V \to \gamma^*$ transition, (b) the diagram for
$V\to e^+e^-$ decay.}
   \label{decay}
\end{figure}

An additional normalization condition consists in relating 
the vector meson wave function to the experimentally measurable
physical quantity --- $V \to e^+e^-$ decay width (Fig.\ref{decay}).
The loop at Fig.\ref{decay}a describes transition $V \to \gamma^*$ and enters
the amplitude of the decay $V \to e^+e^-$ (Fig.\ref{decay}b).
Let us define the decay constant via relation
\begin{equation}
{\cal A} = i\langle 0|J_\mu^{em}|V\rangle = - i f_V c_V \sqrt{4\pi\alpha} V_\mu \,.
\label{b1}
\end{equation}
So defined $f_V$ has dimension dim$[f_V] = m^2$.
The quantity $c_V$ reflects the flavor content of a vector meson
(in the previous calculations it simply gave unity) and is equal to
\begin{equation}
c_V = {1 \over \sqrt{2}}, {1 \over 3\sqrt{2}}, - {1 \over 3}, {2 \over 3}\label{b2}
\end{equation}
for $\rho, \omega, \phi, J/\psi$ mesons correspondingly.

Knowing that such a loop does not mix polarization states, we can multiply both sides of
eq.(\ref{b1}) by $V^*$ and get the expression
\begin{equation}
i f_V = {(-1) \over (2\pi)^4} N_c \cdot \int d^4 k
{ Sp\{i \hat V^* \cdot i(\hat k + m) \cdot i \hat V \Gamma
\cdot i(\hat k - \hat q_V + m)\}
\over (k^2-m^2+i\epsilon) \cdot ((k-q_V)^2 -m^2 +i\epsilon)}\,. \label{b3}
\end{equation}

Calculations similar to the above normalization condition
derivation yield (for the naive type of vertex)
\begin{equation}
f_V = {N_c \over (2\pi)^3} \cdot \int {dz \over z(1-z)} d^2 \vec k
\ \psi_V [-M^2(V V^*) - 4 (V p)(V^* p)]\,.
\label{b4}
\end{equation}

Applying now the projector technique, one gets in the case of $S$ states
(after proper angular averaging)
\begin{equation}
\framebox(200,40){$ \dst f^{(S)} = \fr{N_c}{(2\pi)^3} \cdot
\int d^3 {\bf p}\ \psi_S \fr{8}{3} (M+m)\,$}
\label{b5}
\end{equation}
and in the case of $D$ wave states
\begin{equation}
\framebox(200,40){$ \dst f^{(D)} = \fr{N_c}{(2\pi)^3} \cdot
\int d^3 {\bf p}\ \psi_D \fr{32}{3} \fr{{\bf p}^4}{M+2m}\,.$}
\label{b6}
\end{equation}

Finally, one can write down the expression for the decay width in terms of
$f_V$:
\begin{equation}
\Gamma(V\to e^+e^-) = {1 \over 32 \pi^2 m_V^2} \cdot {m_V \over 2} 4\pi |A|^2 =
{4 \pi \alpha^2 \over 3 m_V^3} \cdot f_V^2 c_V^2\,.\label{b8}
\end{equation}
This formula can be used to extract the numerical value of $f_V$ from experimental data.

\subsection{Ansatz for LCWF}

In order to perform some model evaluations, one should fix a particular form
for the vector meson wave function.
We will give just two examples often used in calculations:
Coulomb--type and oscillator type wave functions.

\subsubsection{Coulomb wave functions}
As one can see from Eqs.(\ref{a6}) and (\ref{a9}),
a LCWF should contain $M$ dependence in form of $M^{-1/2}$
(the normalization condition for non-relativistic wave function
deals with $\int d^3p |\psi_{NR}|^2$~).
So, in this ansatz the normalized wave functions read
\begin{eqnarray}
\psi_{1S}({\bf p}^2) &=& {c_1 \over \sqrt{M}}
{1 \over  ({\bf p}^2+ \alpha_1^2)^2}; \nonumber\\[2mm]
\psi_{2S}({\bf p}^2) &=& {c_2 \over \sqrt{M}}
{({\bf p}^2 - \alpha_2^2) \over ({\bf p}^2+ \alpha_2^2)^3}; \nonumber\\[2mm]
\psi_{D}({\bf p}^2) &=& {c_D \over \sqrt{M}}
{1 \over ({\bf p}^2 + \alpha_D^2)^4} \,.\label{coulomb1}
\end{eqnarray}
with normalization constants to be determined from Eqs.(\ref{a6})
and (\ref{a9}). Here parameters $\alpha_i$ are connected to the
size of a bound system: in the coordinate representation $\psi_i
\propto \exp(- r \alpha_i)$. For usual Coulomb functions $3
\alpha_D = 2 \alpha_2 = \alpha_1 = 1/R_{Bohr}$, where $R_{Bohr}$
is the Bohr radius. However, this relation should be treated with
care in the case of $q\bar{q}$ quarkonia, where 
the quark-antiquark potential is quite complicated and therefore 
$\alpha_i$ should rather be considered as free parameters.

As an example, let's compute the ratio $f_D / f_{1S}$ with these wave functions
in the non-relativistic limit, when $M \approx m_V \approx 2m$.
It will be useful later in the analysis of heavy vector mesons.
\begin{equation}
{f_D\over f_S} =
 \sqrt{80 \over 9} \sqrt{R_S^3 \over R_D^3} \cdot { 1 \over m_V^2 R_D^2}\,.
\label{coulomb3}
\end{equation}
Here $R=1/\alpha$ is a typical size of a vector meson in a given state.

\subsubsection{Oscillator type LCWF}
In this ansatz one has
\begin{eqnarray}
  \psi_{1S} & = & {c_1 \over \sqrt{M}}\exp\left(-{{\bf p}^2 R_1^2 \over 2}\right)\,;
\nonumber\\
  \psi_{2S} & = & {c_2 \over \sqrt{M}}\left(1 - {2 \over 3}{\bf p}^2 R_2^2 \right)
\exp\left(-{{\bf p}^2 R_2^2 \over 2}\right); \nonumber\\
  \psi_{D} & = & {c_D \over \sqrt{M}} \exp\left(-{{\bf p}^2 R_D^2 \over 2}\right)
\,.\label{oscillator1}
\end{eqnarray}
Note again that for purely oscillator potential one also has relation
$R_D = R_1 = R_2$, which might not hold in our case, since
the oscillator type potential is also a crude approximation 
of the true quark-antiquark interaction.
A similar calculation for $f_D/f_S$ ratio yields
\begin{equation}
{f_D\over f_S} =
 \sqrt{160 \over 3} \sqrt{R_S^3 \over R_D^3} \cdot { 1 \over m_V^2 R_D^2}\,.
\label{oscillator3}
\end{equation}
Note that the oscillator ansatz gives at least
$\sqrt{6}$ times larger $f_D/f_S$ ratio.
Therefore, one must admit that such model dependent calculations
are plagued by a great deal of arbitrariness, which
will inevitably affect the results.

\newpage

\section{Vector meson production amplitudes}

\subsection{Preliminary notes}

\begin{figure}[!htb]
   \centering
   \epsfig{file=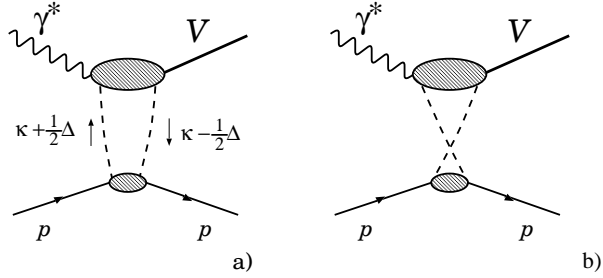,width=80mm}
   \caption{The QCD--inspired diagrams for $\gamma^* p \to Vp$ process
with two gluon $t$--channel. Only Diagr.(a) does contribute to the
imaginary part of amplitudes.}
   \label{main1}
\end{figure}

Having set up the notation and defined and described a vector meson
by itself, we are now ready to evaluate the full set of amplitudes
of its off-forward virtual diffractive photoproduction.

\begin{figure}[!htb]
   \centering
   \epsfig{file=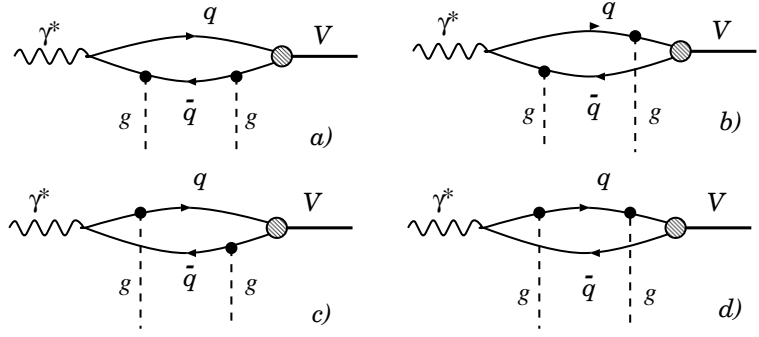,width=100mm}
   \caption{The content of the upper blob in Fig.\ref{main1}a in the
pQCD approach. The true vector meson internal structure is approximated
by $q\bar q$ Fock state.}
   \label{main2}
\end{figure}

In the pQCD motivated approach to this process the pomeron
exchange is viewed as a two-gluon exchange as it is shown in
Fig.\ref{main1}a. Using the scalarization procedure, we will split
the diagram into 2 pieces and will treat each of them separately.
The upper blob describes the pomeron-assisted transition of the
virtual photon into a vector meson. In the perturbative QCD
approach, which is legitimate here due to the presence of the
relevant hard scale $\overline Q^2 = m_q^2 + z(1-z)Q^2$, the
$q\bar q$ fluctuation of the virtual photon interacts with two
hard gluons and then fuses to produce a vector meson. This
interaction is described by four diagrams given in
Fig.\ref{main2}, with all possible two-gluons attachments to $q
\bar q$ pair taken into account. All of them are equally important
and needed for maintaining gauge--invariance and color
transparency. The latter property means that in the case of very
soft gluons the upper blob must yield zero, for the $q \bar q$
pair is colorless.

The lower part of the general diagram Fig.\ref{main1}a
is of course not computable in pQCD. The physically meaningful procedure is
to relate it to the experimentally measurable gluon density.
To do so, we will first calculate this lower blob in the Born approximation
and then give a prescription how to introduce the unintergrated gluon density.
In the course of this procedure, we will neglect in the intermediate
calculations proton off-forwardness  and take it into account only
at the very end, as a certain factor to the unintegrated gluon density.

\begin{figure}[!htb]
   \centering
   \epsfig{file=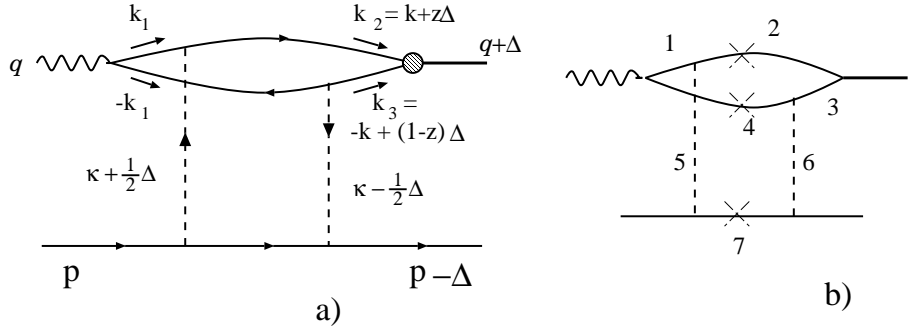,width=120mm}
   \caption{(a): A particularly useful convention of the loop momenta
(only transverse components of quark momenta are shown).
(b): the propagator notation used while calculating denominators.
The crosses denote on mass shell particles.}
   \label{main3}
\end{figure}

\subsection{Notation and helicity amplitudes}

In our calculation we will use the following Sudakov's
decomposition (see also Fig.\ref{main3}a)
\begin{eqnarray}
k_\mu= -y{p_\mu}' + z{q_\mu}' + \vec k_\mu\,;\nonumber\\
\kappa_\mu = \alpha {p_\mu}' - \beta {q_\mu}' + \vec \kappa_\mu\,;\nonumber\\
\Delta_\mu = \delta {p_\mu}' + \sigma {q_\mu}' + \vec \Delta_\mu\label{c1}
\end{eqnarray}
Here $k$ and $\kappa$ are momenta that circulate in the quark 
and gluon loops respectively; $\Delta$ is the momentum transfer. 
Vectors ${p_\mu}'$ and ${q_\mu}'$ denote the light-cone momenta: they are
 related to the proton and virtual photon momenta as
\begin{equation}
p_\mu = {p_\mu}' + {m_p^2 \over s}{q_\mu}';
\quad {q_\mu} = {q_\mu}'-x{p_\mu}';\quad q'^2 = p'^2 = 0;
\quad x = {Q^2 \over s}\ll 1; \quad s = 2(p'q')\,.
\label{c2}
\end{equation}
As it was mentioned in the Introduction, the condition $x\ll 1$ is
necessary to speak about diffractive processes. The longitudinal
momentum transfer can be readily found from kinematics 
(see Fig.\ref{main1}). To the higher power $s$ terms it reads
\begin{eqnarray}
m_p^2 =p^2=(p-\Delta)^2=m_p^2-2(p\Delta) + \Delta^2;
&\Rightarrow &\sigma = - {\vec \Delta^2 \over s}\;;\nonumber\\
m_V^2 = (q +\Delta)^2 = -Q^2 + 2(q\Delta) +\Delta^2
&\Rightarrow&  \delta = x + {m_V^2 + \vec \Delta^2 \over s}\;.
\label{c3}
\end{eqnarray}
The final vector meson momentum reads:
\begin{equation}
q_{V\mu} = {q_\mu}' + {m_V^2 + \vec \Delta^2 \over s}{p_\mu}' + \vec \Delta_\mu\,.
  \label{c2a}
\end{equation}
Finally, throughout the text transverse momenta will be marked by the vector sign
as $\vec k$ and 3D vectors will be written in bold.

There are several possible helicity amplitudes in
the transition $\gamma^*_{\lambda_\gamma} \to V_{\lambda_V}$.
First of all, both photon and vector meson can be transversely polarized.
The polarization vectors are
\begin{equation}
e_{T\mu} = \vec e_\mu\,; \quad
V_{T\mu} = \vec V_\mu + {2 (\vec \Delta \vec V) \over s}(p' -q')_\mu\,.
\label{polart}
\end{equation}
Note that we took into account the fact that
the vector meson momentum has finite transverse component
$\vec \Delta$.
Then, the virtual photon can have the scalar polarization
(which is often called longitudinal;
we will use both terms)
with polarization vector
\begin{equation}
e_{0\mu} = {1 \over Q}(q' + xp')_\mu\,.\label{e0}
\end{equation}
Finally, the longitudinal polarization state of a vector meson
is described by
\begin{equation}
V_{L\mu} = {1 \over M}\left( {q_\mu}' + {\vec \Delta^2 -M^2 \over s}{p_\mu}' 
+ \vec \Delta_\mu \right)\,.\label{VL}
\end{equation}
Note that, as we already mentioned, in the self-consistent approach
we must take the {\it running polarization vector} for 
the longitudinal polarization state. It depends on $M$, not $m_V$, 
which reflects the fact that in our approach we first calculate the
production of {\it an on-shell $q\bar q$ pair with} 
(whose dynamics is governed by $M$) 
and then projects it onto the physical vector meson. We
 stress that this projection will automatically arise 
in the course of usual Feynman diagram evaluation.

Thus, there are 5 different amplitudes:
\begin{eqnarray}
&&L \to L\nonumber\\
&&T \to T\  (\lambda_\gamma = \lambda_V)\nonumber\\
&&T\to L\nonumber\\
&&L\to T\nonumber\\
&&T \to T\  (\lambda_\gamma = -\lambda_V)\label{five}
\end{eqnarray}
The first two are helicity conserving amplitudes.
They are dominant and almost insensitive
to the momentum transfer $\vec \Delta$.
The next two are single helicity flipping amplitudes.
They are unavoidably proportional to $|\vec \Delta|$
in the combination $(\vec e\vec \Delta)$ or $(\vec V^*\vec \Delta)$
and would be vanishing for the strictly forward scattering.
Finally, the last amplitude corresponds to the double helicity flip
and will be proportional to $(\vec e\vec \Delta)(\vec V^*\vec \Delta)$.\\

\subsection{General amplitude}
We will take diagr.(c) at Fig.\ref{main2}
(it is shown in fig.\ref{main3}) as a generic diagram
and perform a thorough analysis for it. It turns out that
the other diagrams are calculated in the same fashion.

The general expression for the amplitude given by diagr.(c) reads:
\begin{eqnarray}
&&iA = \int{d^4k\over (2\pi)^4}\int{d^4\kappa\over (2\pi)^4}
\ \bar{u}_p' (-ig\gamma^{\nu'} t^{B'})
i{\hat p -\hat\kappa_1 +m_p \over \left[(p-\kappa)^2 - m_p^2 +
i\epsilon\right]}(-ig\gamma^{\mu'} t^{A'})u_p
\nonumber\\&&
\cdot(-i){g_{\mu\mu'}\delta_{AA'}\over \kappa_1^2 - \mu^2 +i\epsilon}
\cdot(-i){g_{\nu\nu'}\delta_{BB'}\over \kappa_2^2 - \mu^2 +i\epsilon}
\cdot c_V \cdot \Gamma^*
\nonumber\\
&&\cdot{
Sp\left\{ i e\hat e\  i(\hat k_4 + m)\ (-ig\gamma^\nu t^B)\ i(\hat k_3+m)
i \hat V^*\ i(\hat k_2+m)\ (-ig\gamma^\mu t^A)\ i(\hat k_1+m)
\right\}
\over \left[k_1^2 - m^2 + i\epsilon\right]
\left[k_2^2 - m^2 + i\epsilon\right]
\left[k_3^2 - m^2 + i\epsilon\right]
\left[k_4^2 - m^2 + i\epsilon\right]
}\nonumber\\
\label{gv1}
\end{eqnarray}
Here $c_V$ is the same as in (\ref{b2}) and $\Gamma$ is the
familiar $q \bar q \to V$ vertex function. Note that we introduced
'gluon mass' $\mu$ in gluon propagators to account for confinement
at a phenomenological level.

Let's first calculate the numerator.

\subsection{Color factor}
\begin{figure}[!htb]
   \centering
   \epsfig{file=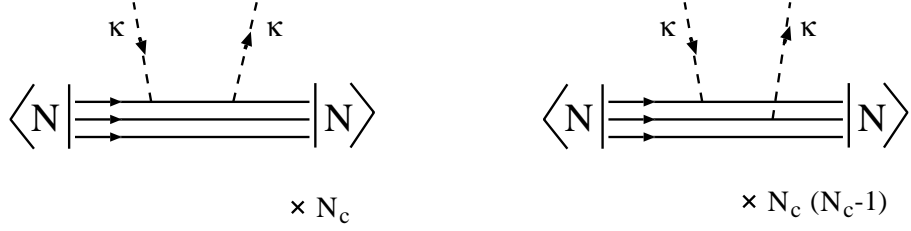,width=120mm}
   \caption{The ways two glons can couple a colorless nucleon.}
   \label{color}
\end{figure}

If we consider {\it strictly forward} gluon scattering off a single
quark, we have
\begin{equation}
{1 \over N_c}Sp\{t^{B'}t^{A'}\} \cdot\delta_{AA'}\delta_{BB'}Sp\{t^Bt^A\}=
{1 \over N_c}{1 \over 2}\delta_{AB}{1 \over 2}\delta_{AB}=
{1 \over 2}{N_c^2-1 \over 2N_c} = {1 \over 2}C_F = {2 \over 3}\label{color1}
\end{equation}
However, we should take into account that quarks are sitting inside
a colorless proton, whose color structure is
\begin{equation}
\psi_{color} = {1 \over \sqrt{6}} \epsilon^{abc} q^a q^b q^c\label{color2}
\end{equation}
In this case there are two ways a pair of gluons can couple 3 quark lines
(see Fig.\ref{color}). In the first way both gluons couple to the same quark.
Since the quark momentum does not change after these two interactions,
the nucleon stays in the same state: $\langle N|N\rangle = 1$.
In the second case gluon legs are
attached to different quark lines, so that extra momentum $\kappa$
circulates between quarks, which gives rise to the factor
 $\langle N|\exp(i\kappa r_1 - i\kappa r_2)|N\rangle$, i.e.
to the two-body formfactor.
Therefore, for the lower line instead of
\begin{equation}
{1 \over N_c}Sp\{t^{B}t^{A}\} = {1 \over N_c} {1 \over 2} \delta_{AB}\label{color3}
\end{equation}
one has
\begin{eqnarray}
&&{1 \over 6} \epsilon^{abc}
\left(3\delta_{aa'}\delta_{bb'} t^A_{cc''} t^B_{c''c'}
+ 6 \delta_{aa'}t^A_{bb'} t^B_{cc'}\langle N|\exp(i\kappa r_1 - i\kappa r_2)|N\rangle
\right)
\epsilon^{a'b'c'}\nonumber\\
&=& Sp\{t^{A}t^{B}\} - Sp\{t^{A}t^{B}\}\langle N|\exp(i\kappa r_1 - i\kappa r_2)|N\rangle
\nonumber\\
& = &{1 \over 2} \delta_{AB}(1 - \langle N|\exp(i\kappa r_1 - i\kappa r_2)|N\rangle
)\,.\label{color4}
\end{eqnarray}
Note also that a similar calculation for $N_c$ number of colors would yield
the same result. Thus, the overall color factor is
\begin{equation}
{1 \over 2} C_F N_c V(\kappa) = 2V(\kappa) \equiv
{1 \over 2} C_F N_c (1 - \langle N|\exp(i\kappa r_1 - i\kappa r_2)|N\rangle).\label{color5}
\end{equation}

\subsection{Scalarization of upper and lower parts}

As known, the highest power $s$ contribution comes from so--called nonsense
components of gluon propagator (density matrix) decomposition:
\begin{equation}
g_{\mu\mu'} = {2 p'_{\mu}q'_{\mu'} \over s} + {2 p'_{\mu}q'_{\mu'} \over s} +
g_{\mu\mu'}^{\bot} \ \approx \ {2 p'_{\mu}q'_{\mu'} \over s}\,.\label{gv2}
\end{equation}
The lower (proton) line gives then
\begin{equation}
\bar u'(p) \cdot \hat q' (\hat p -  \hat \kappa_1 + m_p)
\hat q' \cdot u(p-\Delta)\label{gv2a}
\end{equation}
To the highest power $s$ order, it can be rewritten as
\begin{equation}
\bar u_p' \hat q' \hat p' \hat q' u_p \bigg|_{forward} =
{1 \over 2} Sp\{\hat p' \hat q' \hat p' \hat q'\} = s^2\,.\label{gv3}
\end{equation}
As we mentioned, the effect of off--forwardness (skewedness)
will be taken into account later.
So, combining all factors, one has for numerator of Eq.(\ref{gv1})
\begin{eqnarray}
&&(4\pi\alpha_s)^2\ \sqrt{4\pi\alpha_{em}}\ c_V\cdot 
{1 \over 2} C_F N_c V(\kappa){4 \over s^2} s^2
\cdot Sp\left\{
\hat e\  (\hat k_4 + m)\ \hat q'\ (\hat k_3 + m)\ \hat V^*\
(\hat k_2+m)\ \hat q'\ (\hat k_1 + m)
\right\}\nonumber\\
&=&(4\pi\alpha_s)^2\ \sqrt{4\pi\alpha_{em}}\ c_V\cdot 
2 C_F N_c V(\kappa) \cdot 2s^2\cdot
I^{(c)}(\gamma^* \to V)\,.\label{gv4}
\end{eqnarray}
Note that we factored out $2s^2$ from the trace 
because it will appear later in all trace calculations.
So, the resulting expression for amplitude (\ref{gv1}) looks like
\begin{eqnarray}
A &=& \ \sqrt{4\pi\alpha_{em}}\ 4 C_F N_c s^2 \ c_V\cdot
\int{d^4k\over (2\pi)^4}\int{d^4\kappa\over (2\pi)^4}\cdot
{(4\pi\alpha_s)^2 V(\kappa) \over \left[(p-\kappa_1)^2 - m_p^2 + i\epsilon\right]
\left[\kappa_1^2 - \mu^2 +i\epsilon\right]
\left[\kappa_2^2 - \mu^2 +i\epsilon\right]}
\nonumber\\
&& \cdot\,{ \Gamma^* I^{(c)}(\gamma^* \to V) \over
\left[k_1^2 - m^2 + i\epsilon\right]
\left[k_2^2 - m^2 + i\epsilon\right]
\left[k_3^2 - m^2 + i\epsilon\right]
\left[k_4^2 - m^2 + i\epsilon\right]
}\label{gv5}
\end{eqnarray}
One can now immediately write similar expressions for the other three
diagrams (Fig.\ref{main2} a,b,d). Indeed, they will differ from
Eq.(\ref{gv5}) only by the last line: they will have different
expressions for traces and propagator structures (i.e. the exact
values for $k_i$.).

\subsection{Denominator evaluation}
Now we turn to the calculation of denominators.
As usual, we implement Sudakov's decomposition (\ref{c1})
and make use of relation
$$
d^4k = {1\over 2} s dy\,dz\,d^2\vec k\,.
$$
The physical picture of the way we will do the
resulting integrals is the following.
We are interested only in the imaginary part of these diagrams.
In fact, it can be shown that at the level of accuracy used here
the diagram in Fig.\ref{main1}a gives rise only to the imaginary part
of the amplitude. The real part is given by Fig.\ref{main1}b
and can be readily found from analiticity (so that here is no need
for additional calculations), however it turns out small due to
smallness of pomeron intercept, so we will neglect it in our
subsequent calculations.

The imaginary part is computed by setting three
particles in the $s$-channel cut on mass shell
(this is illustrated in Fig.\ref{main3}b). One way to do so is to apply
Cutkosky rule to modify our expression. Another, more straightforward
and 'honest' way is to calculate three of the integrals
(namely, over $\alpha, \beta, y$) via residues.
That's what we are going to do.

The details of this calculations are given in Appendix \ref{apa}.
Here we cite the result:
\begin{eqnarray}
&&\int dy\ dz\ d\alpha\ d\beta\ {\Gamma \over [\mbox{all propagators}]}
\nonumber\\
&&= \left(-{2\pi i \over s}\right)^3\cdot
\int {dz \over z(1-z)} \psi_V(z,\vec k^2) \cdot{1 \over
[\vec k_1^2 +m^2 +z(1-z)Q^2]} { 1 \over (\vec \kappa^2 + \mu^2)^2}\label{gv14a}
\end{eqnarray}
Here $\vec k_1$ is the transverse momentum flowing through photon vertex
along the fermion line.
Particularly, for diagr.(c) it is equal to 
$\vec k_1 = \vec k - (1-z)\vec \Delta - \vec \kappa_2$
(with the specific quark loop momentum choice given at Fig.(\ref{main3}a)).

Let us now stop for a moment and take a look at this result. As we
know, aside from longitudinal momenta integration, we also have
outer integrations over transverse gluon momenta $\int_0^\infty
d^2\vec \kappa$. It will also be shown later that the upper loop will
always (except for double spin flip amplitude) give factor
$\vec \kappa^2$ in numerator. Thus, we will deal with integral
\begin{equation}\label{de3}
  \int_0^\infty d^2\vec\kappa {\vec\kappa^2 \over (\vec\kappa^2 + \mu^2)^2}
  {1 \over [(\vec\kappa-\vec k+ (1-z)\vec \Delta)^2 + m^2 + z(1-z)Q^2]}\,.
\end{equation}
This integral is naturally split up into three regions: $\vec\kappa^2
< \mu^2$, $\overline Q^2 > \vec\kappa^2 > \mu^2$, and $\vec\kappa^2 >
\overline Q^2$. 
(Here we marked all low energy scales, namely, $\mu^2, m^2, \vec\Delta^2$
by a general parameter $\mu^2$.) A crucial observation here is that the largest
contribution comes from the {\it perturbative} intermediate region
\begin{equation}\label{de4}
  \mu^2 \ll \vec\kappa^2 \ll \overline Q^2
\end{equation}
due to an {\it extra large logarithmic factor} $\log(\overline Q^2
/\mu^2)$ (which of course implies $\overline Q^2 \gg \mu^2$).
{\bf It is precisely the presence of this large logarithm that
justifies our perturbative QCD approach}. Thus, we now established
that in our problem the pQCD approach is valid only in the {\bf
leading log approximation} and has only logarithmic accuracy.\\

Thus, the amplitude for diagr.(c) has the form
\begin{eqnarray}
A &=& \sqrt{4\pi\alpha_{em}}\ 4C_F N_c s^2 \ c_V
\cdot {1 \over 2}s \ {1 \over 2}s \cdot \left(-{2\pi i \over s}\right)^3
\cdot {1 \over (2\pi)^8} \nonumber\\
&&\cdot \int {dz \over z(1-z)} d^2\vec k \psi_V(z,\vec k^2)
\int {d^2 \vec \kappa V(\kappa) \over (\vec \kappa^2 + \mu^2)^2}
\ (4\pi\alpha_s)^2\
\cdot {I^{(c)}\over [\vec k^2_{1} + m^2 + z(1-z)Q^2]}
\,.\label{gv15}
\end{eqnarray}
After bringing all coefficients together, one gets
\begin{eqnarray}
&&A^{(a)} = is {C_F N_c c_V\sqrt{4\pi\alpha_{em}}  \over (2\pi)^5}\cdot
\int {dz \over z(1-z)} d^2\vec k \psi_V(z,\vec k^2)
\int {d^2 \vec \kappa V(\kappa)\over (\vec \kappa^2 + \mu^2)^2}
(4\pi\alpha_s)^2
\nonumber\\
&& \cdot {I^{(c)}\over \vec k^2_{1} + m^2 + z(1-z)Q^2}\,.\label{gv16}
\end{eqnarray}

The other diagrams are calculated in the same way. The most
important difference is that for each diagram we will have a
propagator $1/[\vec k_1^2 + m^2 +z(1-z)Q^2]$ with its own
definition of $\vec{k}_1$, the transverse momentum in photon vertex:
\begin{eqnarray}
&\mbox{diagr.a} & \vec{k}_{1a} = \vec{k} - (1-z)\vec\Delta = \vec{r} - {1\over 2}
\vec\Delta\nonumber\\ &\mbox{diagr.b} & \vec{k}_{1b} = \vec{k} - (1-z)\vec\Delta +
\vec\kappa + {1\over 2} \vec\Delta = \vec{r} + \vec\kappa \nonumber\\
&\mbox{diagr.c} & \vec{k}_{1c} = \vec{k} - (1-z)\vec\Delta - \vec\kappa + {1\over 2}
\vec\Delta = \vec{r} - \vec\kappa \nonumber\\ &\mbox{diagr.d} & \vec{k}_{1d} = \vec{k} +
z\vec\Delta = \vec{r} + {1\over 2} \vec\Delta\label{g1}
\end{eqnarray}
Here $\vec{r} \equiv \vec{k} - (1-2z)\vec\Delta/2$. 
Thus, the whole expression for the imaginary part of the amplitude
is
\begin{eqnarray}
A &=& is {C_F N_c c_V\sqrt{4\pi\alpha_{em}}  \over (2\pi)^5}\cdot
\int {dz \over z(1-z)} d^2\vec k \psi_V(z,\vec k^2)
\int {d^2 \vec \kappa V(\kappa)\over (\vec \kappa^2 + \mu^2)^2}
\,(4\pi\alpha_s)^2\nonumber\\
&&\times \Biggl[
{1-z \over z} {I^{(a)} \over \vec{k}_{1a}^2 + m^2 + z(1-z)Q^2}
 + {I^{(b)} \over \vec{k}_{1b}^2 +m^2 + z(1-z)Q^2}\nonumber\\
&& + {I^{(c)} \over \vec{k}_{1c}^2 +m^2 + z(1-z)Q^2}
+{z \over 1-z}{ I^{(d)} \over \vec{k}_{1d}^2 + m^2 + z(1-z)Q^2}
\Biggr]
\,.\label{gv17}
\end{eqnarray}

\subsection{Final exact results for the naive vertex}

The only thing left to be computed is integrands $I^i(\gamma\to
V)$. For convenience, their calculation is also given in Appendix
\ref{apb}. It turns out that the results can be written in the
same way for all four diagrams via $\vec k_1$ given by (\ref{g1}),
i.e. all quantities:
$$
-{1-z \over z}I^{(a)}\,,\quad I^{(b)}\,,\quad I^{(c)}\,,\quad -{z
\over 1-z}I^{(d)}
$$
can be written in a similar way:
\begin{eqnarray}
T \to T&& \left[(\vec{e}\vec{V}^*)(m^2 + \vec{k}\vec k_1) + (\vec{V}^*\vec{k})(\vec{e}\vec k_1)(1-2z)^2 -
(\vec{e}\vec{k})(\vec{V}^*\vec k_1)\right]\nonumber\\
L \to L&&-4z^2(1-z)^2QM\nonumber\\
T \to L&&2z(1-z)M(\vec{e}\vec k_1)(1-2z)\nonumber\\
L \to T&&-2z(1-z)Q(1-2z)(\vec{V}^*\vec{k})\nonumber
\end{eqnarray}

Therefore, we can cast amplitude (\ref{gv17}) in a compact form
with aid of functions $\vec \Phi_{1}$ and $\Phi_2$:
\begin{equation}
\Phi_2 = -{1 \over (\vec{r}+\vec\kappa)^2 + \overline Q^2} -{1 \over
(\vec{r}-\vec\kappa)^2 + \overline Q^2} + {1 \over (\vec{r} + \vec\Delta/2)^2 +
\overline Q^2} + {1 \over (\vec{r} - \vec\Delta/2)^2 + \overline
Q^2}\label{g2}
\end{equation}
and
\begin{equation}
\vec{\Phi}_1 = -{\vec{r} + \vec\kappa \over (\vec{r}+\vec\kappa)^2 + \overline Q^2}
-{\vec{r} - \vec\kappa \over (\vec{r}-\vec\kappa)^2 + \overline Q^2}
+ {\vec{r} + \vec\Delta/2 \over (\vec{r} + \vec\Delta/2)^2 + \overline Q^2}
+ {\vec{r} - \vec\Delta/2 \over (\vec{r} - \vec\Delta/2)^2 + \overline Q^2}\label{g5}
\end{equation}
With these functions, for the naive $q \bar q V$ vertex the whole
expression in square brackets in (\ref{gv17}) wit sign minus (which we denote
here as $I_{\lambda_\gamma \lambda_V}$) has the form:
\begin{eqnarray}
I_{LL} &=& - 4 QM z^2 (1-z)^2 \Phi_2\,;\nonumber\\
I_{TT} &=& (\vec{e}\vec{V}^*)[m^2\Phi_2 + 
(\vec{k}\vec{\Phi}_1)] + (1-2z)^2(\vec{k}\vec{V}^*)(\vec{e}\vec{\Phi}_1) 
- (\vec{e}\vec{k})(\vec{V}^*\vec{\Phi}_1)\,;\nonumber\\
I_{TL} &=& 2Mz(1-z)(1-2z)(\vec{e}\vec{\Phi}_1)\,;\nonumber\\
I_{LT} &=& -2Qz(1-z)(1-2z)(V\vec{k})\Phi_2\,.\label{f3}
\end{eqnarray}

\subsection{Gluon density}

We have just calculated our reaction in the Born approximation.
We understand however that the nature of $t$-channel exchange
in diffractive processes is much more complicated.
A physically meaningful procedure is to related this
and other diffractive processes to the experimentally
measurable gluon density. Here we explain
a prescription of how to introduce unintegrated gluon
density in answers calculated in the Born approximation.

A well known QED formula for Weizsaker-Williams approximations
unambiguously defines the number of photons in an electron
(in the Born approximation and to the logarithmic accuracy)
\begin{equation}
dn_\gamma = {\alpha_{em} \over 2\pi } [1 + (1-\beta)^2]
{d\vec\kappa^2 \over \vec\kappa^2} {d\beta \over \beta}
\approx {\alpha_{em} \over \pi }
{d\vec\kappa^2 \over \vec\kappa^2} {d\beta \over \beta}\,.\label{density1}
\end{equation}
Here $\beta$ is the fraction of electron energy carried by photon. Quantity
\begin{equation}
\rho_\gamma = \beta{dn_\gamma \over d\beta}
= {\alpha_{em} \over \pi }
{d\vec\kappa^2 \over \vec\kappa^2}\label{density2}
\end{equation}
is called photon $\beta$-density. Thus, the unintegrated photon density is
given by
\begin{equation}
{\cal F}^{(Born)} = { \partial \rho_\gamma \over \partial \log \vec\kappa^2} =
{\alpha_{em} \over \pi}\,.\label{density3}
\end{equation}
In QCD, in the Born approximation, one can define a similar quantity, 
namely, the unintegrated gluon density in a proton,
which after careful accounting for all color factors reads:
\begin{equation}
{\cal F}^{(Born)} = { \partial G \over \partial \log \vec\kappa^2} =
C_F N_c {\alpha_{s} \over \pi} V(\kappa)\,.\label{density4}
\end{equation}
Therefore, a prescription of how to include unintegrated gluon density
is as follows:
\begin{equation}
4 {\alpha_{s} \over \pi} V(\kappa) \equiv {\cal F}^{(Born)} \to
{\cal F}\label{density5}
\end{equation}
Finally, one has now to account for skewedness of this gluon density.
Within diffraction cone (i.e. for $\vec\Delta^2$ up to several GeV$^2$)
it is usually parameterized as
\begin{equation}
{\cal{F}}(x,\vec \kappa,\vec \Delta)=
{\partial G(x,\vec\kappa^{2})\over \partial \log \vec\kappa^{2}}
\exp(-{1\over 2}B_{3\Pom}\vec \Delta^2)\,.
\label{density6}
\end{equation}
where $B_{3\Pom}\sim$ 6 GeV$^{-2}$ is the so-called diffraction
cone slope.

Therefore, a general amplitude of transition
$\gamma^*_{\lambda_\gamma} \to V_{\lambda_V}$ has the following
form
\begin{center}
\framebox(350,80){
\parbox{16cm}{
\begin{eqnarray}
A(x,Q^{2},\vec \Delta)=
- is{c_{V}\sqrt{4\pi\alpha_{em}}
\over 2\pi^{2}}
\int_{0}^{1} {dz\over z(1-z)} \int d^2 \vec k \psi(z,\vec k)
\nonumber\\
\int {d^{2} \vec \kappa
\over
\vec\kappa^{4}}\alpha_{S}{\cal{F}}(x,\vec \kappa,\vec \Delta)
\cdot I(\gamma^{*}\to V)\, ,
\label{f1}
\end{eqnarray}}}
\end{center}

\subsection{Final results for $S$ and $D$ wave amplitudes}

Now we can use projector technique to obtain results for $S$/$D$
wave states.
\begin{eqnarray}
&&V_\mu \to V_\nu {\cal S}_{\nu\mu}\,; \quad
{\cal S}_{\mu\nu} = g_{\mu\nu} - {2p_\mu p_\nu \over m (M+2m)}
\ \Rightarrow\ I^S = I + {2 ({\bf V}{\bf p}) \over m(M+2m)}
p_\mu \otimes \gamma_\mu\,;\label{f4}\\
&&V_\mu \to V_\nu {\cal D}_{\nu\mu}\,; \quad
{\cal D}_{\mu\nu} = {\bf p}^2 g_{\mu\nu} + {(M+m)p_\mu p_\nu \over m}
\ \Rightarrow\ I^D = I{\bf p}^2 -  {(M+m) ({\bf V}{\bf p}) \over m}
p_\mu \otimes \gamma_\mu\,;\nonumber
\end{eqnarray}
Note that $({\bf V}{\bf p})$ is 3D scalar product.
While contracting, we encounter terms proportional to $p_\mu \otimes \gamma_\mu$
which should be understood as
\begin{equation}
p_\mu \otimes \gamma_\mu = I_{V_T}\{\vec V \to \vec p\}
+ I_{V_L}\{1 \equiv V_z \to p_z \equiv {1 \over 2}(2z-1)M\} .\label{f5}
\end{equation}
The result of this substitution reads:\\
for $e_T$:
\begin{eqnarray}
&&I_{T\to T}\{\vec V \to \vec p\}
+ I_{T\to L}\{1 \equiv V_z \to p_z \equiv {1 \over 2}(2z-1)M\} \nonumber\\
&&=m^2 \left[ (\vec{e}\vec{k})\Phi_2 - (\vec{e}\vec{\Phi}_1)(1-2z)^2\right]\label{f6}
\end{eqnarray}
for $e_0$:
\begin{eqnarray}
&&I_{L\to T}\{\vec V \to \vec p\}
+ I_{L\to L}\{1 \equiv V_z \to p_z \equiv {1 \over 2}(2z-1)M\} \nonumber\\
&&=- 2Qz(1-z)(2z-1)m^2 \Phi_2\label{f7}
\end{eqnarray}

So, the resulting integrands for $S$ wave type mesons are
\begin{center}
\framebox(480,200){
\parbox{16cm}{
\begin{eqnarray}
I^S_{L\to L} &=& - 4 QM z^2 (1-z)^2
\left[ 1 + { (1-2z)^2\over 4z(1-z)} {2m \over M+2m}\right] \Phi_2\,;\nonumber\\[1mm]
I^S_{T\to T} &=& (\vec{e}\vec{V}^*)[m^2\Phi_2 + (\vec{k}\vec{\Phi}_1)] + (1-2z)^2(\vec{k}\vec{V}^*)(\vec{e}\vec{\Phi}_1){M \over M+2m}
\nonumber\\&&- (\vec{e}\vec{k})(\vec{V}^*\vec{\Phi}_1) + {2m \over M+2m}(\vec{k}\vec{e})(\vec{k}\vec{V}^*)\Phi_2\,;
\nonumber\\[1mm]
I^S_{T\to L} &=& 2Mz(1-z)(1-2z)(\vec{e}\vec{\Phi}_1)
\left[ 1 + { (1-2z)^2\over 4z(1-z)} {2m \over M+2m}\right]
- {Mm\over M+2m}(1-2z)(\vec{e}\vec{k})\Phi_2\,;\nonumber\\[1mm]
I^S_{L\to T} &=& -2Qz(1-z)(1-2z)(\vec{V}^*\vec{k}){M \over M+2m}\Phi_2\,.\label{f8}
\end{eqnarray}}}
\end{center}
and for $D$ wave type mesons are
\begin{center}
\framebox(400,150){
\parbox{16cm}{
\begin{eqnarray}
I^D_{L\to L} &=& - QM z (1-z)
\left( \vec{k}^2 - {4m \over M}p_z^2 \right)  \Phi_2\,;\nonumber\\[2mm]
I^D_{T\to T} &=& (\vec{e}\vec{V}^*){\bf p^2}[m^2\Phi_2 + (\vec{k}\vec{\Phi}_1)]
+ (1-2z)^2({\bf p}^2 + m^2 + Mm)(\vec{k}\vec{V}^*)(\vec{e}\vec{\Phi}_1) \nonumber\\
&&- {\bf p}^2(\vec{e}\vec{k})(\vec{V}^*\vec{\Phi}_1) - m(M+m)(\vec{k}\vec{e})(\vec{k}\vec{V}^*)\Phi_2\,;
\nonumber\\[2mm]
I^D_{T\to L} &=& {1 \over 2}M(1-2z)
\left[(\vec{e}\vec{\Phi}_1)\left( \vec{k}^2 - {4m \over M}p_z^2 \right)
+ m(M+m)(\vec{e}\vec{k})\Phi_2 \right] \,;\nonumber\\[2mm]
I^D_{L\to T} &=& -2Qz(1-z)(1-2z)(\vec{V}^*\vec{k})({\bf p}^2 + m^2 + Mm)\Phi_2\,.\label{f9}
\end{eqnarray}}}
\end{center}

Equations (\ref{f8}), (\ref{f9}) together with expression
(\ref{f1}) constitute the ultimate sets of all helicity amplitudes.
They give explicit answers for the vector meson production amplitudes
within leading-log-approximation.

\newpage

\section{Analysis for heavy quarkonia}

The general answers (\ref{f8}), (\ref{f9}) are of course incomprehensible
at a quick glance. Therefore, a further analysis is needed to
grasp the most vivid features of the results and to disentangle
$s$-channel helicity conserving and double helicity flip amplitudes.

Since in the heavy vector mesons quarks can be treated
non-relativistically, further simplifications in analytical
formulas (\ref{f8}), (\ref{f9}) are possible due to the presence
of an additional small parameter ${\bf p}^2/m^2$.

In what follows we will first perform the twist expansion and then
relate simplified amplitudes to the decay constants (\ref{b5}),
(\ref{b6}). We will then analyze twist hierarchy of the amplitudes
and compare results for $S$ vs. $D$ wave states. Though we perform
this analysis for heavy mesons, we wish to stress that all
qualitative features ($S$ vs. $D$ difference, $Q^2$ dependence
etc.) will hold for light quarkonia as well.

\subsection{Twist expansion}
Here we are going to expand the amplitudes (or to be more exact,
the quantities $\vec \Phi_1$ and $\Phi_2$ (\ref{g2}), (\ref{g5}))
in inverse powers of the hard scale $\overline Q^2$ and
then perform azimuthal angular averaging over $\phi_\kappa$.

Expanding $\Phi_2$ in twists in the main
logarithmic region
\begin{equation}
\mu^2, \vec\Delta^2, \vec{k}^2 \ll \vec\kappa^2 \ll \overline Q^2\,,
\label{g3}
\end{equation}
one observes that
twist--1 terms cancel, so one has to retain twist--2 and twist--3 terms
proportional $\vec\kappa^2$:
\begin{equation}
\Phi_2 =
 {2 \vec\kappa^2 \over \overline Q^4} - {8 \vec\kappa^2 \vec{r}^2 \over \overline Q^6}\label{g4}
\end{equation}

The analogous decomposition for $\vec \Phi_{1}$ reads
\begin{equation}
\vec \Phi_{1} =
{4 \vec r \vec\kappa^2 \over \overline Q^4} - {12\vec r
\vec\kappa^2 r^2 \over \overline Q^6}
 - {\vec \Delta (\vec{r}\vec\Delta)\over \overline Q^4} \label{g6}
\end{equation}
Note that the last term does not contain $\vec\kappa^2$.
However, one must track it because it will be important
in double helicity flip amplitudes.

\subsection{Twist expansion for $S$ wave type mesons}

With the aid of this decomposition one obtains:\\
for amplitude $L\to L$
\begin{equation}
I^S_{L\to L} = -4QMz^2(1-z)^2{2 \vec\kappa^2 \over \overline Q^4}
\left[ 1 + {(1-2z)^2\over 4z(1-z)} {2m \over M+2m}\right]\,,\label{h3}
\end{equation}
for amplitude $T\to T$
\begin{eqnarray}
&&I^S_{T\to T} = (\vec{e}\vec{V}^*)\left[m^2{2 \vec\kappa^2 \over \overline Q^4}
+{4 \vec\kappa^2 \over \overline Q^4} \vec{k}^2\right]
\ +\ {2m \over M+2m}\cdot {1 \over 2}\vec{k}^2 (\vec{e}\vec{V}^*){2 \vec\kappa^2 \over \overline Q^4}
\label{h4}\\
&&+ \left[(1-2z)^2{M \over M+2m} - 1\right]
\left[ \vec{k}^2 (\vec{e}\vec{V}^*){2 \vec\kappa^2 \over \overline Q^4} -
{\vec{k}^2 \over 2\overline Q^4}(\vec{e}\vec\Delta)(\vec{V}^*\vec\Delta)
\left(1 + {6 \vec\kappa^2 (1-2z)^2 \over \overline Q^2}\right) \right] \nonumber
\end{eqnarray}
This amplitude is naturally split into $s$--channel helicity conserving
and double helicity flip parts
\begin{eqnarray}
I^S_{T\to T}(\lambda_\gamma = \lambda_V) &= &
(\vec{e}\vec{V}^*){2 \vec\kappa^2 \over \overline Q^4}
\left[ m^2 + 2 \vec{k}^2(z^2 + (1-z)^2) + {m\over M+2m} \vec{k}^2 (1-2(1-2z)^2)\right]
\nonumber\\
I^S_{T\to T}(\lambda_\gamma = - \lambda_V) &= &
4z(1-z)(\vec{e}\vec\Delta)(\vec{V}^*\vec\Delta){\vec{k}^2 \over 2\overline Q^4}
\left(1 + {6 \vec\kappa^2 (1-2z)^2 \over \overline Q^2}\right)
\left[ 1 + {(1-2z)^2\over 4z(1-z)} {2m \over M+2m}\right]
\nonumber\\ \label{h5}
\end{eqnarray}
Finally, single spin flip amplitudes are
\begin{eqnarray}
I^S_{T\to L} &=& - 2Mz(1-z)(1-2z)^2 {2 \vec\kappa^2 \over \overline Q^4}(\vec{e}\vec\Delta)
\left[ 1 + {(1-2z)^2\over 4z(1-z)} {2m \over M+2m}\right]\,,\nonumber\\
I^S_{L\to T} &=&-2 Qz(1-z)(1-2z)^2{2 \vec\kappa^2 \over \overline Q^4}(\vec{V}^*\vec\Delta)
{2 \vec{k}^2 \over \overline Q^2}{M\over M+2m}\,.\label{h6}
\end{eqnarray}

\subsection{Twist expansion for $D$--type vector mesons}
Here we will need to track higher--twist terms.
It will turn out later that leading contributions vanish,
so twist--3 terms will be crucial for our results.\\
For amplitude $L\to L$ one has
\begin{equation}
I^D_{L\to L}=-QMz(1-z)\left(\vec{k}^2 - {4m\over M}p_z^2\right)
\cdot {2 \vec\kappa^2 \over \overline Q^4} \left(1 - {4 \vec{k}^2\over \overline Q^2}\right)\,.
\label{h7}
\end{equation}
For $T\to T$ amplitude, one obtains
\begin{eqnarray}
&&I^D_{T\to T}\ =\ (\vec{e}\vec{V}^*){\bf p}^2 \left[m^2{2 \vec\kappa^2 \over \overline Q^4}
\left(1 - {4 \vec{k}^2\over \overline Q^2}\right)
+ {2 \vec\kappa^2 \over \overline Q^4} 2\vec{k}^2  \right]\nonumber\\
&&+\left[-4z(1-z){\bf p}^2 + (1-2z)^2m(M+m)\right]
\left\{ {2 \vec\kappa^2 \over \overline Q^4} \vec{k}^2 (\vec{e}\vec{V}^*) -
{\vec{k}^2 \over 2\overline Q^4}(\vec{e}\vec\Delta)(\vec{V}^*\vec\Delta)
\left[1 + {6 \vec\kappa^2(1-2z)^2 \over \overline Q^2}\right] \right\}\nonumber\\
&&-m(M+m) {1 \over 2}\vec{k}^2 (\vec{e}\vec{V}^*){2 \vec\kappa^2 \over \overline Q^4}
 \left( 1 - {4 \vec{k}^2\over \overline Q^2} \right)\label{h8}
\end{eqnarray}
Note that we kept track of all terms $\propto |p|^4$. Again, one
can separate out $s$-channel helicity conserving and double
helicity flip parts:
\begin{eqnarray}
I^D_{T\to T}(\lambda_\gamma = \lambda_V) &= &
(\vec{e}\vec{V}^*){\vec\kappa^2 \over \overline Q^4}
\Biggr[ 2{\bf p}^2\left( m^2 + 2\vec{k}^2 - 4\vec{k}^2 {m^2 \over \overline Q^2}\right)
-m(M+m)\vec{k}^2 \left(1 - {4 \vec{k}^2 \over \overline Q^2}\right)
\nonumber\\
&&-2\vec{k}^2 \left(\vec{k}^2 - {4m \over M}p_z^2\right)\Biggr]
\nonumber\\
I^D_{T\to T}(\lambda_\gamma = - \lambda_V) &= &
\left(\vec{k}^2 - {4m \over M}p_z^2\right)
(\vec{e}\vec\Delta)(\vec{V}^*\vec\Delta){\vec{k}^2 \over 2\overline Q^4}
\left(1 + {6 \vec\kappa^2 (1-2z)^2 \over \overline Q^2}\right)
\nonumber\\ \label{h9}
\end{eqnarray}

Finally, single helicity flipping amplitudes are
\begin{eqnarray}
I^D_{T\to L} &=& {1 \over 2}M(1-2z)
\left[{-2(1-2z)(\vec{e}\vec\Delta)\vec\kappa^2 \over \overline Q^4}\left(\vec{k}^2 - {4m\over M}p_z^2\right)
+ m(M+m){4\vec\kappa^2 \vec{k}^2 \over \overline Q^6}(1-2z)(\vec{e}\vec\Delta)
\right]\nonumber\\
&=&- {\vec\kappa^2 \over \overline Q^4}(1-2z)^2M(\vec{e}\vec\Delta)
\left[\vec{k}^2 - {4m\over M}p_z^2 - m(M+m){2 \vec{k}^2 \over \overline Q^2}\right]\,,\nonumber\\
I^D_{L\to T} &=& - 8Qz(1-z)(1-2z)^2({\bf p}^2 + m^2 + mM)
{\vec\kappa^2 \over \overline Q^6}\vec{k}^2 (\vec{V}^*\vec\Delta)\,.\label{h10}
\end{eqnarray}

\subsection{Final results for $S$ wave mesons}

In order to grasp the major features of various $S$ and $D$ wave
amplitudes, further simplifications can be achieved if one
neglects spherically non-symmetric arguments of $\alpha_s$ and
gluon density. First we rewrite general expression (\ref{f1}) in
the more convenient form
\begin{eqnarray}
A(x,Q^{2},\vec \Delta)=
-is{c_{V}\sqrt{4\pi\alpha_{em}}
\over 2\pi^{2}}
\int d^3 {\bf p} {4 \over M} \psi({\bf p}^2)\int {d^{2} \vec \kappa \over \kappa^{4}}
\alpha_{S}{\cal{F}}\cdot I(\gamma^{*}\to V)\, .
\label{i1}
\end{eqnarray}
In this expression everything except for integrands $I(\gamma^{*}\to V)$ is
spherically symmetric, thus making it possible to perform angular
averaging over $\Omega_{\bf p}$ in these integrands.

\subsubsection{$S$ wave: $\Omega_{\bf p}$ averaging}

Here all the calculations are rather straightforward.
In the non-relativistic case one can everywhere put $z \to 1/2\,; M =2m=m_V$.
The resultant integrands are:
\begin{center}
\framebox(370,200){
\parbox{16cm}{
\begin{eqnarray}
I^S(L\to L) &=& - {8QM \over (Q^2 + M^2)^2} \vec\kappa^2\nonumber\\[2mm]
I^S(T \to T; \lambda_\gamma = \lambda_V) &=& {8 M^2 \over (Q^2 + M^2)^2} \vec\kappa^2
\nonumber\\[2mm]
I^S(T \to T; \lambda_\gamma = -\lambda_V) &=&
{16 \over 3} {(\vec{e}\vec\Delta)(\vec{V}^*\vec\Delta) \over (Q^2 + M^2)^2}
\left[ 1 + {96 \over 5} {\vec\kappa^2 {\bf p}^2 \over M^2 (Q^2
+M^2)}\right]\nonumber\\[2mm]
I^S(T\to L) &=& - {64 \over 3} {M (\vec{e}\vec\Delta) \over (Q^2 +M^2)^2}
{{\bf p}^2 \over M^2} \vec\kappa^2\nonumber\\[2mm]
I^S(L\to T) &=& - {512 \over 15} {Q(\vec{V}^*\vec\Delta) \over (Q^2+M^2)^2}
{{\bf p}^4 \over M^2 (Q^2 +M^2)}\vec\kappa^2
\label{i3}
\end{eqnarray}}}
\end{center}
Note several things: since the accurate $1S$ wave differs from the naive $\gamma_\mu$
spinorial structure only by relativistic corrections,
one would obtain {\it the same results} in the case of naive $q\bar q V$ vertex.
The only difference would be only extra factor 2 for the $L\to T$ amplitude
(which is higher-twist amplitude, anyway).

Thus the only thing left is $|{\bf p}|$ integration.
Note that these amplitudes are naturally expressed in terms of decay constants.
Indeed, in the extremely non-relativistic case expression (\ref{b5})
turns into
\begin{equation}
f^{(S)} = {3m_V \over 2\pi^3} \int d^3{\bf p}\ \psi_S\quad \Rightarrow\quad
\int d^3{\bf p}\ \psi_S = {2\pi^3 \over 3m_V} f^{(S)}
\label{i4}
\end{equation}

\subsubsection{$S$ wave: answers for $L \to L$ up to differential
cross section}

Here we would like to digress and for the sake of logical
completeness show how one obtains the final result for the
differential cross section with the example of $L\to L$ amplitude. If
needed, the same can be done for the other amplitudes, so we will
do it just once.

One has:
\begin{eqnarray}
&&\int d^3{\bf p}\ \psi_S  {4 \over m_V} {-8Qm_V \over (Q^2 +m_V^2)^2}
\cdot \pi \int^{\overline Q^2} {d\vec\kappa^2 \over \vec\kappa^2}
{\partial G(x,\vec\kappa^{2})\over \partial \log \vec\kappa^{2}}
\alpha_s(\vec\kappa^2)\exp(-{1\over 2}B_{3\Pom}\vec \Delta^2) \nonumber\\
&&=- {32 \pi Q \over (Q^2+m_V^2)^2}
G(x,\overline Q_0^2) \alpha_s(\overline Q_0^2)\exp(-{1\over 2}B_{3\Pom}\vec \Delta^2)
\int d^3{\bf p}\ \psi_S\nonumber\\
&&=- {32 \pi Q \over (Q^2+m_V^2)^2} \exp(-{1\over 2}B_{3\Pom}\vec \Delta^2)
G(x,\overline Q_0^2) \alpha_s(\overline Q_0^2)
{2\pi^3 \over 3m_V} f_V\nonumber
\end{eqnarray}
With this result, (\ref{i1}) becomes
\begin{eqnarray}
A &=& i s {f_V c_V \sqrt{4\pi\alpha_{em}} \over 2\pi^2}
\cdot {64 \pi^4 \over 3} {Q \over m_V} {G\cdot \alpha_s \over  (Q^2 +m_V^2)^2}
\nonumber\\
 &=& is{32 \pi^2 \over 3} {Q \over m_V} \cdot c_V f_V \cdot \sqrt{4\pi\alpha_{em}}
 {G\cdot \alpha_s \exp(-{1\over 2}B_{3\Pom}\vec \Delta^2)\over  (Q^2 +m_V^2)^2}
\nonumber
\end{eqnarray}
The expression for the differential cross section reads ($t \equiv \vec \Delta^2$):
\begin{eqnarray}
{d\sigma \over dt}&=& {1 \over 16\pi s^2}|A|^2\nonumber\\
&=&{64 \pi^3 \over 9} {Q^2 \over m_V^2} \cdot (c_V f_V)^2 4\pi\alpha_{em}
\cdot {G^2 \alpha_s^2  \exp(-B_{3\Pom}t) \over (Q^2 +m_V^2)^4}\,.
\nonumber
\end{eqnarray}
Finally, one can express this cross section through $\Gamma(V \to e^+e^-)$
(see (\ref{b8})):
\begin{equation}
\framebox(330,50){$ \dst
{d\sigma \over dt}={64 \pi^3 \over 3 \alpha_{em}} Q^2 m_V \cdot\Gamma(V \to e^+e^-)
\cdot {G^2 \alpha_s^2  \exp(-B_{3\Pom}t) \over (Q^2 +m_V^2)^4}\,.$}\label{i8}
\end{equation}

\subsubsection{$S$ wave: the other amplitudes}
{\bf $T \to T, \lambda_\gamma = \lambda_V$}.\\
In the non-relativistic case, this amplitude is readily obtained from the above formulas
after $Q \to m_V$ replacement in the numerator of the amplitude (see (\ref{i3})).
This means in particular that in this limit
\begin{equation}
R^{(S)} \equiv \left({A_{LL}\over A_{TT}}\right)^2
= {Q^2 \over m_V^2}\,.\label{i10}
\end{equation}

{\bf $T \to T, \lambda_\gamma = -\lambda_V$}.\\
This amplitude is very interesting because of the competition of two
very different terms --- soft and hard scale contribution. Indeed,
integration over gluon loop gives
\begin{equation}
A \propto {G(x,\mu^2) \over \mu^2} +
{96 \over 5} {G(x,\overline Q^2) {\bf p}^2 \over M^2 (Q^2 +M^2)}\label{i11}
\end{equation}
We see that the soft contribution turns out to be
of leading twist, while the pQCD contribution is of higher twist.
This observation was first made in \cite{IK} for the naive type
of $q \bar q V$ vertex; here we see that it also holds for
accurate $S$ and $D$ wave vector mesons.

Since at moderate $Q^2$ both terms can be comparable, we cannot
neglect soft scale contribution, which means that we failed to
compute the double helicity flip amplitude within pQCD approach.
\\

{\bf $T \to L$ and $L \to T$}.\\
In the case of heavy quarkonia
these single spin flip amplitudes have suppressing
non-relativistic factors. Besides, the amplitude $L \to T$ is of
twist 3, which is another source of suppression. Their ratios to
$A(T\to T) \equiv A(T\to T; \lambda_\gamma = \lambda_V^*)$ read

\begin{equation}
{A(T\to L) \over A(T\to T)} = - {8 \over 3} {(\vec{e}\vec\Delta) \over m_V} \cdot w_2\,;\quad
{A(L\to T) \over A(T\to T)}
= - {64 \over 15} {Q(\vec{V}^*\vec\Delta) \over Q^2 +  m_V^2} \cdot w_4\,.\label{i12}
\end{equation}
The model dependent quantities $w_2$ and $w_4$ are defined via
\begin{equation}
w_2 = {1 \over m_V^2}
{\int d^3{\bf p} \ {\bf p}^2\ \psi_S \over \int d^3{\bf p} \ \psi_S}\,;\quad
w_4 = {1 \over m_V^4}
{\int d^3{\bf p} \ {\bf p}^4\ \psi_S \over \int d^3{\bf p} \ \psi_S}\,.\label{i13}
\end{equation}
Within the oscillator ansatz (\ref{oscillator1}) their values are
\begin{equation}
w_2 = {3 \over 2}{1 \over (m_V R)^2}\,;\quad
w_4 = {15 \over 4}{1 \over (m_V R)^4}\,.\label{i14}
\end{equation}

\subsection{Final results for $D$ wave}

This case is much more tricky. It turns out that the leading terms
in integrands $I^{(D)}$, proportional to $m^2 |p|^2$ cancel out
after angular averaging, so that many new terms, including higher
twist terms come into play. This cancellation is in fact quite
understandable. Indeed, in the very beginning we showed that
vertex $\bar u' \gamma_\mu u$ contains both $S$ and $D$ waves,
with $D$ wave probability being suppressed for heavy quarks due to
non-relativistic motion. This means in particular that the {\it
photon} couples to $q \bar q$ pairs sitting either in $S$ or $D$
wave state. However, at the other end of the quark loop we have a
vector meson in pure $D$ wave state. Therefore, the largest items
in $\langle \gamma_{S+D}|...|V_D\rangle$ cancel out due to
$S$--$D$ orthogonality.

\subsubsection{$D$ wave: $\Omega_{\bf p}$ averaging for $L\to L$ amplitude}

If we limited ourselves only to the leading ${\bf p}^2/m^2$ terms,
we would get
$$
\int d\Omega_{\bf p} \left(\vec{k}^2 - {4m\over M}p_z^2\right)
= 4\pi\cdot \left({2\over 3}{\bf p}^2 - 2\cdot{1\over 3}{\bf
p}^2\right)= 0\,,
$$
which is the manifestation of $S$--$D$ orthogonality.
Thus, we see that ${\bf p}^2/m^2$ terms vanish after angular
averaging. Therefore, one has to be extremely careful now
and must take into account all possible sources of
${\bf p}^4/m^4$ terms.
To do so, one has to perform the following averaging
\begin{equation}
\left\langle 4z(1-z)\cdot{1 \over \overline Q^4}\cdot
\left(\vec{k}^2 - {4m\over M}p_z^2\right)\cdot \left(1 - {4 \vec{k}^2\over \overline Q^2}
\right)\right\rangle
\label{i15}
\end{equation}
The detailed evaluation given in Appendix \ref{apc} results in
$$
{1 \over \overline Q_0^4} {4 {\bf p}^4 \over 15M^2}\left(1 - 8{M^2
\over Q^2 + M^2}\right)\,,
$$
where $\overline Q_0^2 = m^2 + Q^2/4$.
One can now again express the integral over quark loop through
the decay constant (see (\ref{b6})):
\begin{equation}
\int d^3{\bf p}\ {\bf p}^4\ \psi_D \Rightarrow f^{(D)}\cdot {\pi^3 m_V \over 2}\,.
\label{i20}
\end{equation}
to give
\begin{equation}
 - {64 \pi^4 \over 15}\; {Q \over m_V} \;
{ G\cdot \alpha_s\cdot \exp(-{1\over 2}B_{3\Pom}\vec \Delta^2)
\over (Q^2+m_V^2)^2}\cdot f^{(D)}\cdot
\left( 1 - 8 {m_V^2 \over Q^2 + m_V^2}\right)\label{i21}
\end{equation}
Comparison with $L\to L$ amplitude reveals that
\begin{equation}
\framebox(220,50){$\dst{A^D_{LL}\over A^S_{LL}} = {1 \over 5}\left( 1 - 8 {m_V^2 \over Q^2 + m_V^2}\right)
\cdot {f^{(D)} \over f^{(S)}}\,.$}\label{i22}
\end{equation}

\subsubsection{$D$ wave: the other amplitudes}

For the helicity conserving amplitude one has to repeat the same
averaging procedure, which is again fully elaborated in Appendix
\ref{apc}. The result can be written as
\begin{equation}
\framebox(180,70){$ \dst { A^D_{LL} \over A^D_{TT}} = {1\over 15}
{ 1 - 8\fr{m_V^2}{Q^2 + m_V^2} \over 1 + \fr{4}{15}
\fr{m_V^2}{Q^2 + m_V^2}}\,.$} \label{i24}
\end{equation}

In the case of double helicity flip amplitude we again have
contributions from soft and hard scales with the same hierarchy
of twists. Namely,
\begin{equation}
A \propto {G(x,\mu^2) \over \mu^2} -
{96 \over 7} {G(x,\overline Q^2) {\bf p}^2 \over M^2 (Q^2 +M^2)}\label{i24a}
\end{equation}
so we again conclude that for the case of double helicity flip amplitude
soft contribution invalidates our pQCD analysis.

In the case of single spin flip amplitudes, no dangerous calcellations among
leading terms arise. Before giving a list of amplitudes, we wish to emphasize that
in the case of $D$ wave mesons there is no non-relativistic suppressing factors
like $w_2$ and $w_4$ defined in (\ref{i13}). This means that for
moderate momentum transfers helicity non-conserving amplitudes are absolutely
important for that case of $D$ wave mesons.\\

\subsection{$S$ wave vs. $D$ wave comparison}

We would like to present our final results in the form which
stresses the remarkable differences between $S$ wave and $D$ wave amplitudes.
Below we give a table of the ratios
\begin{equation}
\rho_{ij} \equiv {A^D(i\to j) \over A^S(i\to j)}\, {f^{(S)} \over f^{(D)}}\label{rho}
\end{equation}
for helicity conserving and single spin flipping amplitudes.
Double spin flip amplitudes are not given due to the presence
of incalculable non-perturbative contributions.
\begin{center}
\framebox(230,150){
\parbox{16cm}{
\begin{eqnarray}
\rho_{LL} &=&{1 \over 5}
\left(1 - 8{m_V^2 \over Q^2 + m_V^2}\right)\nonumber\\
\rho_{TT} &=& 3
\left(1 + {4 \over 15}{m_V^2 \over Q^2 + m_V^2}\right)\nonumber\\
\rho_{TL} &=&
- {3 \over 5} {1 \over w_2} \left(1 + 3{m_V^2 \over Q^2 + m_V^2}\right)\nonumber\\
\rho_{LT} &=&
{9 \over 8}{1 \over w_4}\label{i26}
\end{eqnarray}}}
\end{center}

Thus, we note several things. First, the abnormally large higher
twist contributions to $D$ wave amplitudes are seen here as terms
$\propto m_V^2/(Q^2 + m_V^2)$. They even force the opposite sign
of $L \to L$ amplitude in the moderate $Q^2$ domain. Second, we
see highly non--trivial and even non--monotonous $Q^2$ dependence
of $(A_{LL}/A_{TT})^2$ ratio, which will lead to the presence of a dip
in experimentally measured $\sigma_L/\sigma_T$ for $D$ wave meson
production. Finally, we must stress that in the case of $D$ wave
mesons there is no non-relativistic suppression for single spin
flip amplitudes as it was in $S$ wave mesons. This leads us to a
conclusion that $s$-channel helicity is strongly violated in the
case of $D$ wave meson production.

\newpage
\begin{figure}[!htb]
   \centering
   \epsfig{file=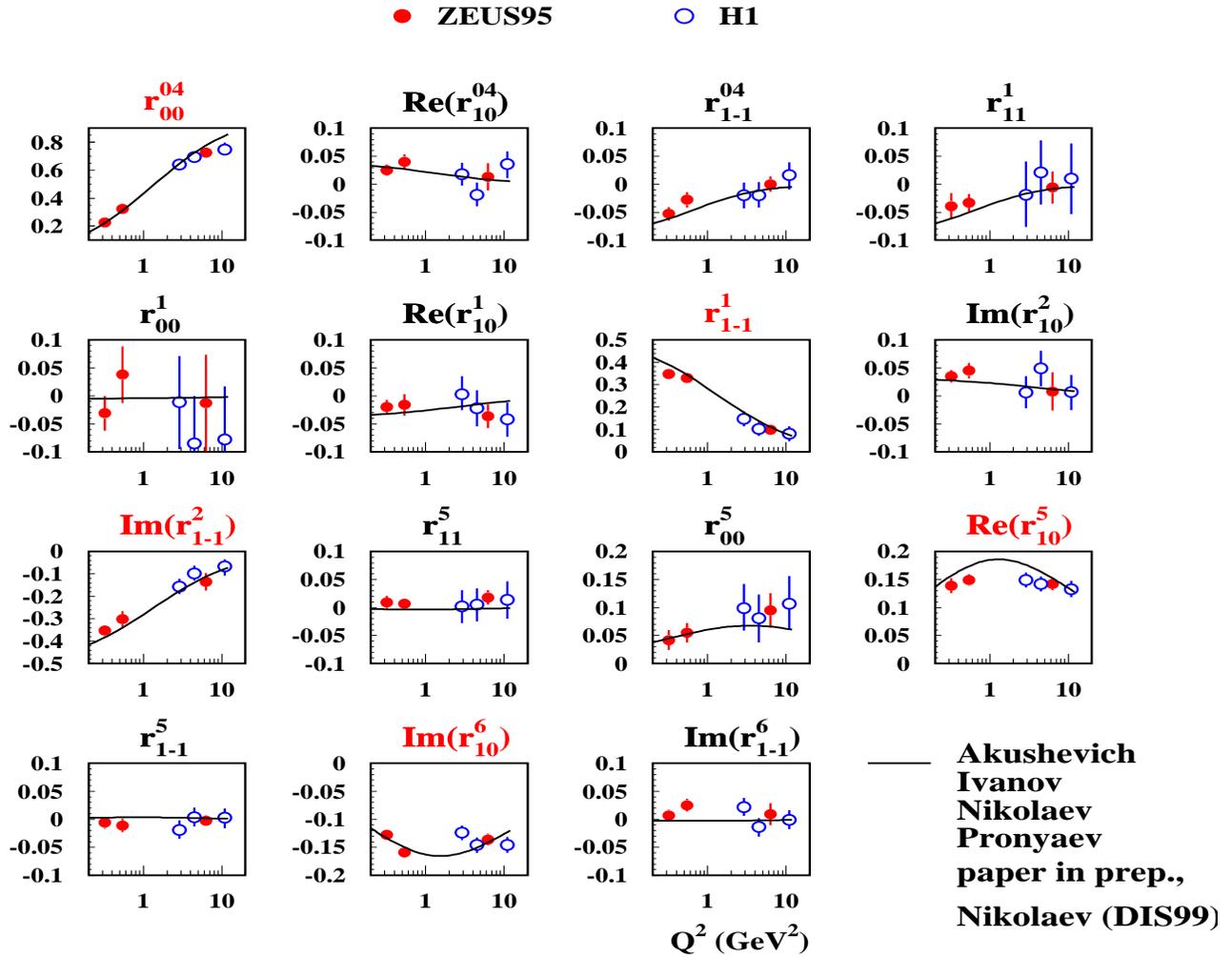,width=170mm}
   \caption{The $\rho$ meson spin density matrix: comparison of
recent HERA data with our predictions for $1S$ state.}
   \label{rho0}
\end{figure}

\section{Numerical analysis}

In this section we will present some results of the numerical
investigation. These results bear mostly illustrative purpose.
Undoubtedly, they do not exhaust all interesting issues
and leave much room for further numerical analysis.

\begin{figure}[!htb]
   \centering
   \epsfig{file=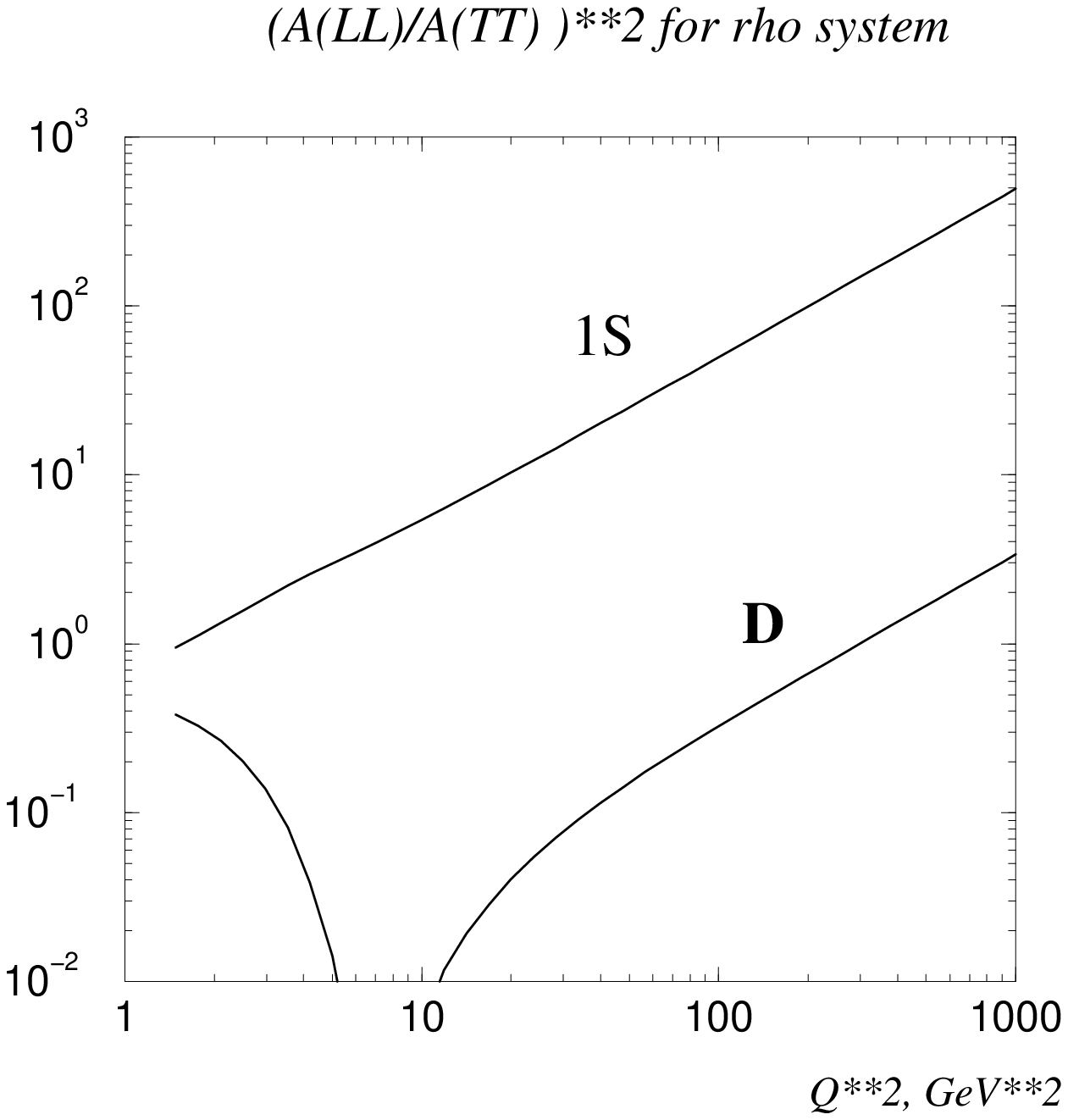,width=80mm}
   \caption{Ratio $[A(L\to L)/A(T \to T)]^2$ for $S$ and $D$ states in $\rho$
system at fixed $x=0.01$ as a function of $Q^2$.}
   \label{rho1}
\end{figure}

Everywhere in our analysis we used oscillator type wave functions
(\ref{oscillator1}) with $R_D = R_{1S}$.
There are several parameters to vary
in order to achive the most correct description of the vector mesons.
Aside from the very ansatz for the wave function, which is now fixed,
we also have quark mass $m$ and radius $R$ as free parameters.
We stress that these are purely phenomenological parameters;
in particular, it would be a mistake to assign values of several MeV
to the quark masses. In principle, these quark masses can even vary
from the ground to excited states for the same quarkonium.

We used normalization conditions (\ref{a6}) and (\ref{a9})
to determine normalization constants
and chose a specific pair $m$, $R$ to have the decay constant
coinciding with the experimental values.
To avoid unnecessarily many free parameters, we used the same
values for $S$ and $D$ wave states.
Specifically, for the $\rho$ meson we took $m = 0.3$ GeV and
$R = 4.0$ GeV$^{-1}$ (which corresponds 0.8 fm).
For charmonium the value were  $m = 1.3$ GeV and
$R = 1.5$ GeV$^{-1} = 0.3$ fm.

Because all quantities displayed in figures are ratios
of one amplitude to another, these results are not sensitive
to the particular gluon distribution parameterization.
Since we leave double helicity flip amplitudes aside,
we can now switch to the integrated gluon density
$$
G(x,\overline Q^2)\exp(B_{3\Pom}\vec\Delta^2) =
\int {d\vec\kappa^2 \over \vec\kappa^2}{\cal F}\,.
$$
So, we used the integrated form of GRV parameterization
with a linear extrapolation of gluon density in the region
$Q^2 < 1$ GeV$^2$ so that $G(x,Q^2) \to 0$ when $Q^2 \to 0$.
In fact, in an accurate numerical analysis
low $Q^2$ region should be treated more carefully.
Indeed, parton distributions do not evolve in this region,
however, they can be parameterized with aid of low energy experimental
data.
\\

\begin{figure}[!htb]
   \centering
   \epsfig{file=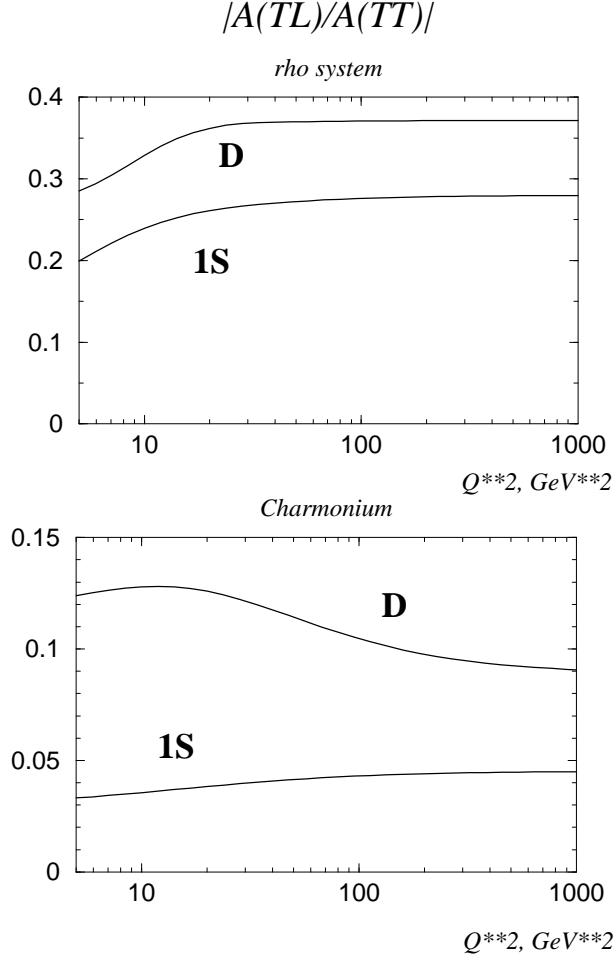,width=80mm}
   \caption{The absolute value of ratio $A(T \to L)/A(T\to T)$ revealing
the relative magnitude of $T\to L$ spin flipping amplitude
at fixed $x=0.01$ as a function of $Q^2$.
The sign of this ratio is $-$ for $S$ wave and $+$ for $D$ wave states.}
   \label{TLTT}
\end{figure}

\begin{figure}[!htb]
   \centering
   \epsfig{file=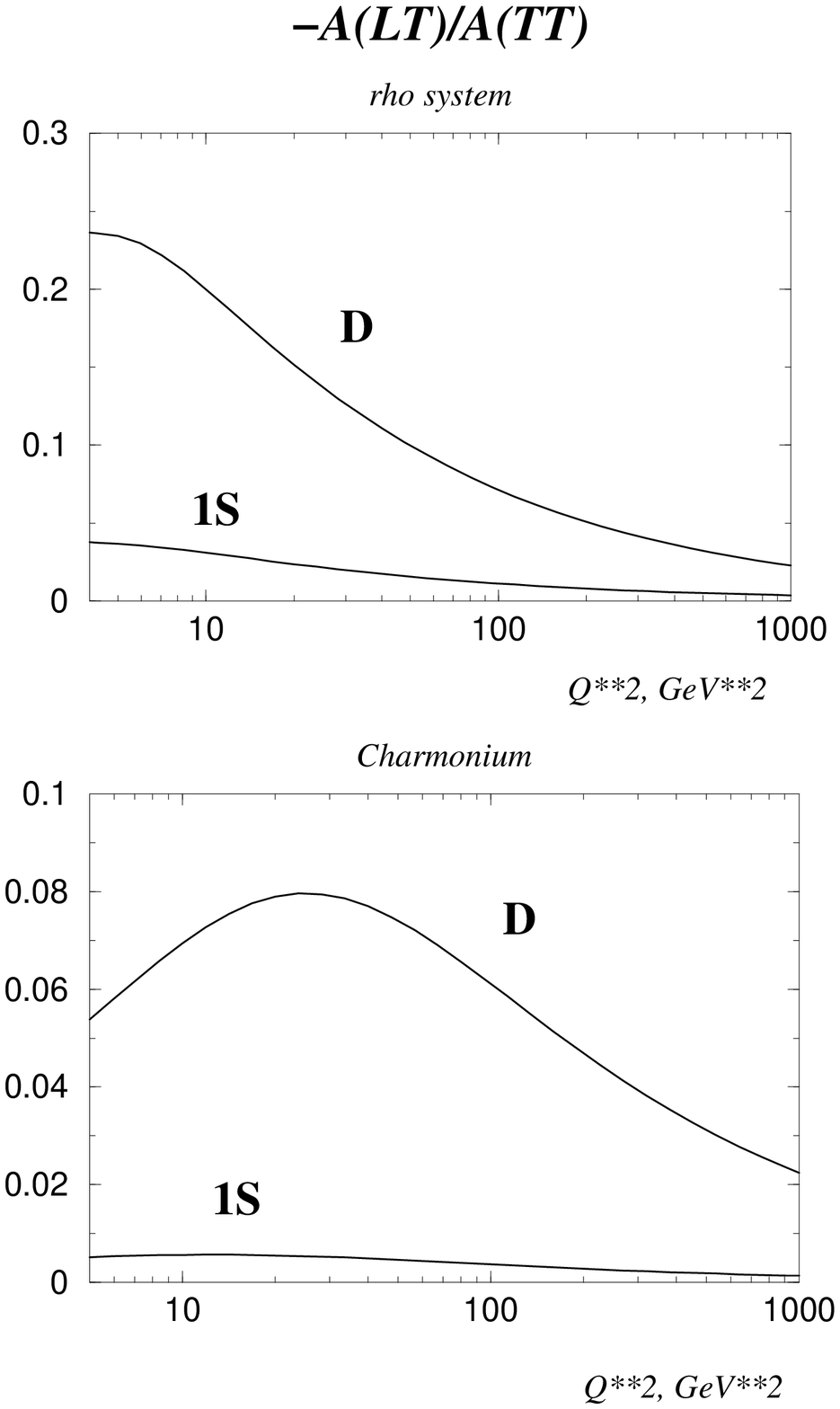,width=80mm}
   \caption{The ratio $-A(L \to T)/A(T\to T)$ showing
the relative magnitude of $L\to T$ spin flipping amplitude
at fixed $x=0.01$ as a function of $Q^2$. It is seen that
this transition is of higher twist.}
   \label{LTTT}
\end{figure}

In Fig.\ref{rho0} we compare our calculations for $\rho$ vector meson
density matrix with recent experimental data from HERA \cite{HERArho}.
We see that our calculations describe data reasonable well,
including parameters connected with $s$-channel helicity violation.\\
\begin{figure}[!htb]
   \centering
   \epsfig{file=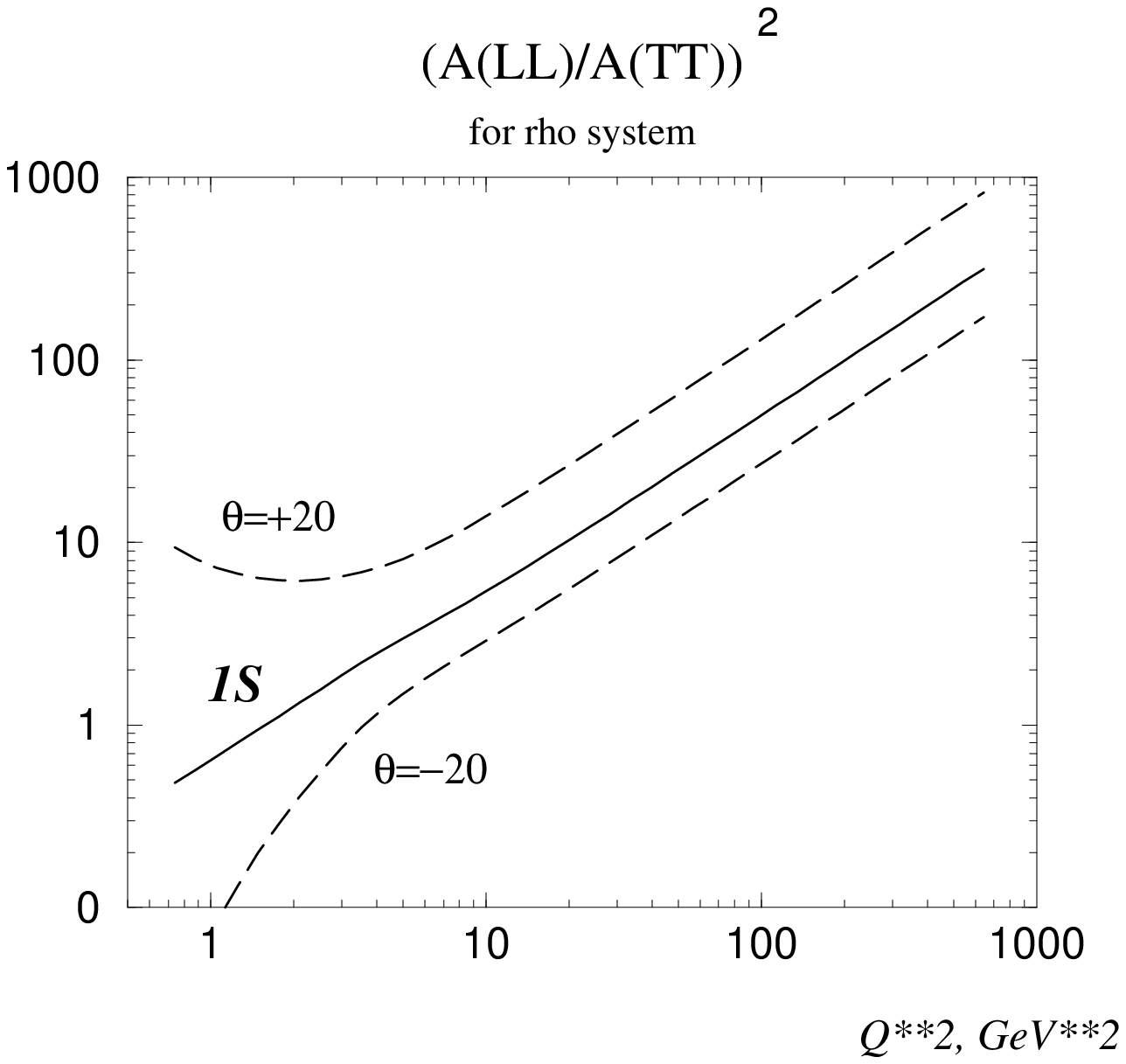,width=100mm}
   \caption{The impact of possible $S-D$ mixing in the physical $\rho$
meson on the ratio $[A(L\to L)/A(T \to T)]^2$ at fixed $x=0.01$
as a function of $Q^2$.}
   \label{mixing}
\end{figure}

In Fig.\ref{rho1} we show ratio $(A_{LL}/A_{TT})^2$
(which would be $\sigma_L/\sigma_T$ if the $s$-channel helicity were conserved)
for both $S$ and $D$ wave states of $\rho$ and $J/\psi$ system.
A striking difference is observed both at small and large $Q^2$.\\

In Figs.\ref{TLTT}, \ref{LTTT}
we present the relative magnitude of helicity non-conserving amplitudes
again for $S$ and $D$ states in $\rho$ and $J/\psi$ system.
It is interesting to note that $T \to L$ amplitudes have opposite signs
for $S$ and $D$ wave mesons.
We underline the large helicity violating amplitudes in $D$ wave states,
both in non-relativistic $J/\psi$ and fully relativistic $\rho$ meson.
The size of the effect reaches 35\% for $\rho$ system and
13\% for charmonium. Thus, it becomes now quite
obvious that in the case of $D$ wave mesons no helicity conservation
holds even at the qualitative level, in accordance with our previous
conclusion from the heavy quarkonia analysis.\\

Finally, we analyzed the possibility of $S-D$ mixing in the physical $\rho$
meson. Namely, we introduced mixing angle $\theta$ as
\begin{equation}
\rho_{phys} = \cos\theta |1S\rangle + \sin\theta |D\rangle \label{mix1}
\end{equation}
and plotted ratio $(A_{LL}/A_{TT})^2$  in Fig.\ref{mixing}
for mixing angles $\theta =\pm 20^\circ$.
The results are strikingly sensitive to the presence of $D$ wave admixture
in the physical $\rho$ meson;
further investigation of this phenomenon is needed.\\

\newpage

\section{Conclusion}

In this work we brought under an accurate analysis the issue of
internal vector meson spin structure in its diffractive production
in DIS. In fact, we developed an essentially new way to study the internal 
spin structure of a vector meson.

To do so, we elaborated a formalism of treating pure $S$ and $D$
wave vector mesons in such processes. For the first time, the full
sets of helicity amplitudes in both cases were calculated. The
exact results (\ref{f1}), (\ref{f8}), (\ref{f9}) were derived
within pQCD leading log approximation and then analyzed
analytically in the limiting case of heavy vector mesons
(\ref{i8}), (\ref{i10}), (\ref{i12}), (\ref{i26}) and numerically
in the case of $\rho$ system and charmonium (Figs. 10--14).

In particular, for the $s$-channel helicity conserving amplitudes
we observed:
\begin{itemize}
  \item remarkably different $Q^2$-dependence of $S$ and $D$ wave
  type amplitudes, providing thus a way to discern $S$ and
  $D$ wave states that are indistinguishable at $e^+e^-$
  colliders;
  \item dramatic role of higher twist contributions to the $D$
  wave vector mesons, which even forced sign change for $L\to L$
  amplitude and led to non-monotonous $\sigma_L/\sigma_T$ ratio;
  \item in the case of light vector mesons we found that $D$ wave
  amplitudes are absolutely comparable with $S$ wave, making,
  for example, the physical $\rho$ meson very sensitive
  to the possible presence of $D$ wave admixture.
\end{itemize}
Our results for helicity violating amplitudes include:
\begin{itemize}
  \item observation that they are not negligible for light vector
  mesons of $S$ wave type, but become suppressed for heavy $S$
  wave mesons;
  \item very large helicity violating effects for $D$ wave vector
  states, which do not get suppressed even in the case of heavy
  quarkonia;
  \item opposite signs for the largest spin flip amplitude
  $T \to L$ for $S$ and $D$ wave vector mesons;
  \item confirmation of the soft dominance of the double spin flip
  amplitude in the case of accurate $S$ and $D$ wave states.
\end{itemize}

Though we performed some numerical analysis, there is much room
for further investigation. In particular, more extensive
analysis is needed to check the model dependence of the numerical
results obtained. It is also desirable to give definite numerical
predictions for diffractive DIS experiments.\\

\newpage

\appendix
\section{Denominator evaluation: details} \label{apa}

Below we give a detailed derivation of Eq.(\ref{gv14a}). The major
guideline will be again analysis of pole positions and 
setting some of the propagators on mass shell by taking
appropriate residues.\\

We first start with the case of strictly forward scattering. 
(Though it corresponds, of course, to forward elastic 
$\gamma^*\to \gamma^*$, not $\gamma^*\to V$ scattering,
it will not have any effect on the expressions we derive here.)
So, the integral to be calculated (see Fig.\ref{main3}) is
\begin{eqnarray}
&&\int dy\ dz\ d\alpha\ d\beta\ {1 \over [(k-\kappa)^2 -m^2
+i\epsilon] [k^2 -m^2 +i\epsilon] [(k-q)^2 -m^2 +i\epsilon]
[(k-q-\kappa)^2 -m^2 +i\epsilon]}\nonumber\\ &&{1 \over [\kappa^2
-\mu^2 +i\epsilon]^2[(p-\kappa)^2 -m_p^2 +i\epsilon]}
\label{gv14b}
\end{eqnarray}
With Sudakov's decomposition (\ref{c1}), one can rewrite all 7
propagators as:
\begin{eqnarray}
&\langle 1\rangle&\quad (k-\kappa)^2 - m^2 +i\epsilon = (-y-\alpha)(z+\beta)s
-((\vec{k}-\vec\kappa)^2 +m^2 -i\epsilon) \,,\nonumber\\ 
&\langle 2\rangle&\quad k^2 -
m^2 +i\epsilon = -szy -(\vec k^2 +m^2 -i\epsilon) \,,\nonumber\\
&\langle 3\rangle&\quad (k-q)^2 - m^2 +i\epsilon = (-y+x)(z-1)s -(\vec{k}^2 +m^2
-i\epsilon)\,,\label{gv9}\\ 
&\langle 4\rangle&\quad (k-q-\kappa)^2 - m^2
+i\epsilon = (-y+x-\alpha)(z-1+\beta)s -((\vec{k}-\vec\kappa)^2 +m^2
-i\epsilon)\,, \nonumber\\ 
&\langle 5,6\rangle&\quad \kappa^2 - \mu^2
+i\epsilon = -\alpha\beta s -(\vec\kappa^2 +\mu^2 -i\epsilon)
\,,\nonumber\\ 
&\langle 7\rangle&\quad (p-\kappa)^2 - m_p^2 +i\epsilon =
(1-\alpha)\beta s -(\vec\kappa^2 +m_p^2 -i\epsilon) \,.\nonumber
\end{eqnarray}
After an accurate analysis of position of all poles on complex
$\alpha, \beta, y$ planes, one finds that the only non-zero
contribution survives if one uses line $\langle 4\rangle$ to extract $\alpha$,
line $\langle 2\rangle$ to extract $y$ and 
line $\langle 7\rangle$ to extract $\beta$:
\begin{equation}
\alpha = - y + x + {(\vec{k}-\vec\kappa)^2 +m^2 \over (1-z)s};
\quad y = - {\vec{k}^2 + m^2 \over zs }; 
\quad \beta = {\vec\kappa^2 + \mu^2\over s}\,.
\label{de1}
\end{equation}
Note that resultant $\alpha, \beta, y \ll 1$. 
The remaining $z$-integration goes from 0 to 1. 
Therefore, integral (\ref{gv14b}) turns into
\begin{equation}\label{gv14c}
\left(-{2\pi i \over s}\right)^3\cdot \int_0^1 {dz \over z}{1
\over 1-z}\cdot {1 \over \langle 5\rangle
\langle 6\rangle \langle 1\rangle \langle 3\rangle}\,,
\end{equation}
where the numbers in brackets denote the corresponding
propagators. With aid of (\ref{de1}), these propagators become:
\begin{eqnarray}\label{de2}
  &\langle 5,6\rangle&\quad -\alpha\beta s - (\vec\kappa^2 + \mu^2) \approx
  - (\vec\kappa^2 + \mu^2)\nonumber\\
  &\langle 3\rangle&\quad y(1-z)s -xs(1-z) -(\vec{k}^2 + m^2) 
= -{1 \over z}[\vec{k}^2 + m^2 +  z(1-z)Q^2]\nonumber\\
  &\langle 1\rangle&\quad -(y+\alpha-x)zs - [(\vec{k}-\vec\kappa)^2 + m^2] =
  -{1 \over 1-z}[(\vec{k}-\vec\kappa)^2 + m^2 + z(1-z)Q^2]\,.
\end{eqnarray}
Here we used $\alpha\ll 1$. Substituting these expressions into (\ref{gv14c}), 
one gets
\begin{equation}\label{gv14d}
\left(-{2\pi i \over s}\right)^3\cdot \int_0^1 {dz \over
(\vec\kappa^2+\mu^2)^2}{1 \over [\vec{k}^2 + m^2 +
  z(1-z)Q^2][(\vec{k}-\vec\kappa)^2 + m^2 + z(1-z)Q^2]}\,.
\end{equation}
Note that we replaced the lower integration limit by zero, knowing
that there is no large end-point contribution.\\

In the above analysis we took momentum transfer $\vec\Delta =0$ for
simplicity. Now let us see what non-zero momentum transfer will
change (see Fig.(\ref{main3})).

First of all, now $\vec\Delta$ will appear in gluon propagators.
However, condition (\ref{de4}) allows us to neglect it within
diffraction cone. Then, $\vec\Delta$ will enter quark loop
propagators. Here the presence of momentum transfer (especially,
the longitudinal one) is very important. Indeed, within our
notation of momenta, the propagator pair $\langle 1,4\rangle$ will yield
the same contribution as before $$ \vec k_1^2 + m^2 + z(1-z)Q^2\,,$$
where $\vec k_1$ is transverse momentum flowing through the photon
vertex. On the other hand, propagators $\langle 2,3\rangle$ will give now
$$ \vec{k}^2 + m^2 - z(1-z)m_V^2\,,$$ which together with vertex factor
$\Gamma$ gives rise to the vector meson LCWF. Combining everything
together, one now has

\begin{eqnarray} &&\int dy\ dz\ d\alpha\ d\beta\ {\Gamma \over
[\mbox{all propagators}]} \nonumber\\ &&= \left(-{2\pi i \over
s}\right)^3\cdot \int {dz \over z(1-z)} \psi(z,\vec{k}^2) \cdot{1 \over
\vec k_1^2 +m^2 +z(1-z)Q^2}\ {1\over (\vec\kappa^2 + \mu^2)^2}\label{gv14}
\end{eqnarray}

\newpage

\section{Helicity amplitude technique evaluation} \label{apb}

Here we give the derivation of expression for traces
of the following type
\begin{equation}
Sp\left\{
\hat e\  (\hat k_4 + m)\ \hat p'\ (\hat k_3 + m)\ \hat V^*\
(\hat k_2+m)\ \hat p'\ (\hat k_1 + m)
\right\}
\label{sp1}
\end{equation}
in full detail. Though one can calculate this trace covariantly,
a particularly convenient way to do so is given by
light cone helicity amplitude technique \cite{LCQFT}.
We emphasize that both ways are absolutely equivalent.
In the helicity amplitude approach, we recall that all
fermion lines in (\ref{sp1}) can be taken on mass shell
(see detailed derivation of LCWF normalization in
Sect.\ref{sectnorm}) and decomposed into a sum of
light cone helicities
\begin{eqnarray}
&&(\hat k + m) \to \sum_{\lambda = \pm} u_\lambda \bar u_\lambda
\quad \mbox{for quark lines;}\nonumber\\
&&(\hat k + m) = - [ (-\hat k) -m ] \to
- \sum_{\lambda = \pm} v_\lambda \bar v_\lambda
\quad \mbox{for antiquark lines.}\nonumber
\end{eqnarray}
Since the specific choice of this decomposition does not
affect the final result, we are free to take the most
convenient choice of spinors (see \cite{LCQFT} for details), namely,
\begin{eqnarray}
u(p,\lambda) &=& {1 \over \sqrt{\sqrt{2}p^+}}\left(\sqrt{2}p^+ + \beta m
+ \vec\alpha\vec p\right) \left\{
\begin{array}{cc}
\chi(\uparrow) & \lambda = +1 \\
\chi(\downarrow) & \lambda = -1 
\end{array}\right. \nonumber\\
v(p,\lambda) &=& {1 \over \sqrt{\sqrt{2}p^+}}\left(\sqrt{2}p^+ - \beta m
+ \vec\alpha\vec p\right) \left\{
\begin{array}{cc}
\chi(\downarrow) & \lambda = +1 \\
\chi(\uparrow) & \lambda = -1 
\end{array}\right.
\label{spinors1a}
\end{eqnarray}
where
$$
\chi(\uparrow) = {1 \over \sqrt{2}}\left(
\begin{array}{c}
1\\ 0\\ 1\\ 0
\end{array}
\right)\,;\quad 
\chi(\downarrow) = {1 \over \sqrt{2}}\left(
\begin{array}{c}
0\\ 1\\ 0\\ -1
\end{array}
\right)
$$
It is a straightforward exercise to show that
they obey usual normalization and completeness requirements:
\begin{eqnarray}
  \label{spinors4}
  &&\bar u(p,\lambda)u(p,\lambda') = - \bar v(p,\lambda)v(p,\lambda') = 
2m \delta_{\lambda\lambda'}\,;\nonumber\\&&
\sum_\lambda u(p,\lambda)\bar u(p,\lambda) = \hat p +m\,;\quad
\sum_\lambda v(p,\lambda)\bar v(p,\lambda) = \hat p -m\,.
\end{eqnarray}

Thus, expression (\ref{sp1}) turns into
\begin{equation}
2s^2\cdot I^{(c)}\ = \ (-1)^2 \sum_{\lambda_i} \bar u \hat e v \cdot
\bar v \hat p' v \cdot \bar v \hat V^* u \cdot
\bar u \hat p' u \cdot \,.
\end{equation}
Though this expression is written for Diagr.\ref{main2}c,
one can immediately redo the same transformation for the other diagrams.
So, in the following we first list the light cone
helicity amplitudes to be used later. Then, we derive the
amplitudes for photon and vector meson vertices $\bar u \hat e v$
and $\bar v \hat V^* u$ respectively with appropriate
polarization vectors. Finally, we give a simple rule
of how to include 2 gluon legs attachments (i.e. $\bar u \hat p' u$
and $\bar v \hat p' v$) and write the final expressions.

\subsection{List of helicity amplitudes}

\begin{figure}[!htb]
   \centering
   \epsfig{file=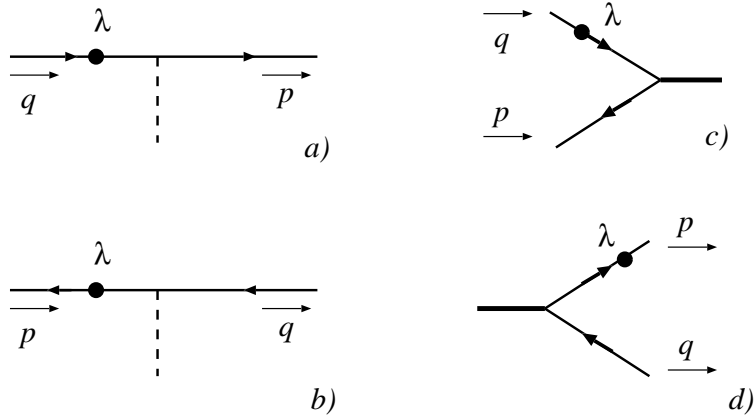,width=100mm}
   \caption{Four types of transitions for which
the light cone helicity amplitudes are given below.
The dot with label $\lambda$ indicates the spinor, whose
helicity is used as $\lambda$ in the table of amplitudes.}
   \label{helicity}
\end{figure}

Here we list various amplitudes of $\bar u'_{\lambda_1}(p)...u_{\lambda_2}(q)$
type various gamma matrix structures inserted between
light cone spinors. 
This notation should not be confused with our previous usage of $p$ and $q$;
here and only here these momenta are assigned to the fermion lines as shown 
at Fig.\ref{helicity}. 

For a derivation of these amplitudes
see \cite{LCQFT}.

Notation: \\
as usual, each vector $\vec a$ should be understood
as $\vec a_\bot$,\\
momenta $p$ and $q$ are constituent momenta shown at relevant figures;\\
For $\lambda= \pm 1$ we define
\begin{equation}
a(\lambda) = - \lambda a_x - i a_y \ .
  \label{eq3}
\end{equation}
'Cross product' is defined as
\begin{equation}
[\vec{a}\vec{b}] = a_x b_y - b_x a_y \ .
  \label{eq4}
\end{equation}
(This is not a true cross product of two vectors, this is just a handy notation.)
Moreover {\bf every} matrix element should be multiplied by common factor
$\sqrt{p_+q_+}$.

\subsubsection{Straight line elements}
Fig.\ref{helicity}a:
\begin{eqnarray}
 & \bar{u}_\lambda(p)...u_\lambda(q) & \bar{u}_{-\lambda}(p)...u_\lambda(q)
\nonumber \\[4mm]
\gamma^+ \quad & 2 & 0 \nonumber\\
\gamma^-  \quad& \fr{1}{p_+q_+}\left( m^2 + \vec{p}\vec{q} + i\lambda [\vec{p}\vec{q}]\right)
& {m \over p_+q_+}\left(p(\lambda) - q(\lambda)\right)  \nonumber\\
\vec{a}\cdot \vec{\gamma} \quad
& \fr{\vec{a}\vec{p}}{p_+} + \fr{\vec{a}\vec{q}}{q_+}
- i \lambda \left( \fr{[\vec{a}\vec{p}]}{p_+} - \fr{[\vec{a}\vec{q}]}{q_+}  \right)
& -m a(\lambda) \left({1 \over p_+} - {1 \over q_+} \right)
\end{eqnarray}
\noindent Fig.\ref{helicity}b
\begin{eqnarray}
& \bar{v}_\lambda(p)...v_\lambda(q) & \bar{v}_{\lambda}(p)...v_{-\lambda}(q)  \nonumber\\[4mm]
\gamma^+ \quad & 2 & 0 \nonumber\\
\gamma^- \quad
& \fr{1}{p_+q_+}\left( m^2 + \vec{p}\vec{q} - i\lambda [\vec{p}\vec{q}]\right)
& - {m \over p_+q_+}\left(p(\lambda) - q(\lambda)\right)   \nonumber\\
\vec{a}\cdot \vec{\gamma}  \quad
& \fr{\vec{a}\vec{p}}{p_+} + \fr{\vec{a}\vec{q}}{q_+}
+ i \lambda \left( \fr{[\vec{a}\vec{p}]}{p_+} - \fr{[\vec{a}\vec{q}]}{q_+}  \right)
& m a(\lambda) \left({1 \over p_+} - {1 \over q_+} \right)
\end{eqnarray}

\subsubsection{Vertex lines}
Fig.\ref{helicity}c
\begin{eqnarray}
 & \bar{v}_\lambda(p)...u_\lambda(q) & \bar{v}_{-\lambda}(p)...u_\lambda(q)
\nonumber \\[4mm]
\gamma^+ \quad & 0 & 2 \nonumber\\
\gamma^-  \quad
& \fr{m}{p_+q_+}\left(p(\lambda) + q(\lambda)\right)
& {1 \over p_+q_+}\left( - m^2 + \vec{p}\vec{q} + i\lambda [\vec{p}\vec{q}]\right)
\nonumber\\
\vec{a}\cdot \vec{\gamma} \quad
& m a(\lambda) \left(\fr{1}{p_+} + \fr{1}{q_+} \right)
& {\vec{a}\vec{p} \over p_+} + {\vec{a}\vec{q} \over q_+}
- i \lambda \left( {[\vec{a}\vec{p}] \over p_+} - {[\vec{a}\vec{q}] \over q_+}  \right)
\nonumber
\end{eqnarray}
\noindent Fig.\ref{helicity}d
\begin{eqnarray}
& \bar{u}_\lambda(p)...v_\lambda(q) & \bar{u}_{\lambda}(p)...v_{-\lambda}(q)  \nonumber\\[4mm]
\gamma^+ \quad & 0 & 2 \nonumber\\
\gamma^-  \quad
& - \fr{m}{p_+q_+}\left(p(-\lambda) + q(-\lambda)\right)
& {1 \over p_+q_+}\left( - m^2 + \vec{p}\vec{q} + i\lambda [\vec{p}\vec{q}]\right)
\nonumber\\
\vec{a}\cdot \vec{\gamma} \quad
& - m a(-\lambda) \left(\fr{1}{p_+} + \fr{1}{q_+} \right)
& {\vec{a}\vec{p} \over p_+} + {\vec{a}\vec{q} \over q_+}
- i \lambda \left( {[\vec{a}\vec{p}] \over p_+} - {[\vec{a}\vec{q}] \over q_+}  \right)
\nonumber
\end{eqnarray}

\subsection{Photon vertex amplitudes}

Notation is given in Fig.\ref{main3}.
We start with transverse photon.
$$
\bar u' \hat e_T v = \bar u' ( - \vec \gamma \vec e) v\,.
$$
Equal $q\bar q$ helicities give
\begin{equation}
- { \sqrt{z(1-z)} q_+ \over z(1-z)q_+}\cdot
(-m) e(-\lambda) = {1 \over \sqrt{z(1-z)}}m e(-\lambda)\label{amp1}
\end{equation}
Opposite $q\bar q$ helicities give
\begin{eqnarray}
&&- { \sqrt{z(1-z)} q_+ \over z(1-z)q_+}\cdot
\left[(\vec{e}\vec k_1)(1-z) - (\vec{e}\vec k_1)z - i\lambda
\left([\vec{e}\vec k_1](1-z) + [\vec{e}\vec k_1]z\right)
\right]\nonumber\\
&&= - {1 \over \sqrt{z(1-z)}}\left[(\vec{e}\vec k_1)(1-2z)- i\lambda[\vec{e}\vec k_1]\right]
\label{amp2}
\end{eqnarray}

In the case of scalar photon
$$
\bar u' \hat e_0 v = \bar u' {1 \over Q}(q_+\gamma_- + xp_-\gamma_+ ) v
$$
the same helicities give exactly zero while opposite helicities
result in
\begin{eqnarray}
&&{ \sqrt{z(1-z)} q_+ \over z(1-z)q_+}\cdot{1 \over Q}\cdot
\left[{q_+ \over q_+} (-m^2 -\vec k_1^2) + x 2z(1-z)p_-q_+\right]\nonumber\\
&&= - {1 \over \sqrt{z(1-z)}}{1 \over Q}
\left[m^2 + \vec k_1^2 - z(1-z)Q^2\right]
\label{amp3}
\end{eqnarray}

\subsection{Vector meson vertex amplitudes}
This case is more tricky due to the nonzero vector meson
transverse momentum $\vec \Delta$.
We start with the  transverse vector meson polarization:
$$
\hat V^*_T = - \vec \gamma \vec V^* + {2 (\vec V^*\vec \Delta) \over s}
p_-\gamma_+\,.
$$
The same $q\bar q$ helicities give again
\begin{equation}
- { \sqrt{z(1-z)} q_+ \over z(1-z)q_+}\cdot
m V^*(\lambda) = -{1 \over \sqrt{z(1-z)}}m V^*(\lambda)\label{amp4}
\end{equation}
while opposite helicities give
\begin{eqnarray}
&&{ \sqrt{z(1-z)} q_+ \over z(1-z)q_+}
\Biggl\{-\bigl[(\vec{V}^*\vec{k}_3)z + (\vec{V}^*\vec{k}_2)(1-z)
- i\lambda\left([\vec{V}^*\vec{k}_3]z - [\vec{V}^*\vec{k}_2](1-z)\right)\bigr]\nonumber\\
&&+ {2 (\vec{V}^*\vec\Delta) \over s}p_-2z(1-z)q_+
\Biggr\} \nonumber\\
 &&= - {1 \over \sqrt{z(1-z)}}\left\{
(\vec{V}^*\vec{k})(1-2z) + i\lambda[V\vec{k}]\right\}\label{amp5}
\end{eqnarray}
Here we used definitions and properties (see also Fig.\ref{main3}):
\begin{eqnarray}
&&\vec{k}_2 = \vec{k} + z\vec\Delta\,; \quad \vec{k}_3 = -\vec{k}+(1-z)\vec\Delta \label{d12}\\
&\Rightarrow& (1-z) \vec{k}_2 - z \vec{k}_3 = \vec{k}\,; \quad
(1-z) \vec{k}_2 + z \vec{k}_3 = (1-2z)\vec{k} + 2 z(1-z) \vec\Delta\,;\nonumber\\ 
&&(\vec{k}_2\vec{k}_3) = -\vec{k}^2 + (1-2z)(\vec{k}\vec\Delta) + z(1-z) \vec\Delta^2\;
\nonumber\\
&&M^2 + \vec \Delta^2 = {\vec k_2^2 +m^2 \over z} +{\vec k_3^2 +m^2 \over (1-z)} =
 {\vec k^2 +m^2 \over z(1-z)} + \vec \Delta^2\,.
\nonumber
\end{eqnarray}

For the longitudinal vector mesons one has
for equal quark-antiquark helicities
\begin{equation}
{ \sqrt{z(1-z)} q_+ \over z(1-z)q_+}{1\over M}
\cdot\left[-m\vec\Delta(\lambda) + m{q_+\over q_+} [k_2(\lambda)+k_3(\lambda)]
\right] = 0\label{amp6}
\end{equation}
and for opposite helicities
\begin{eqnarray}
&&{ \sqrt{z(1-z)} q_+ \over z(1-z)q_+}{1\over M} \Biggl\{
-\bigl[(\vec\Delta \vec{k}_3)z + (\vec\Delta \vec{k}_2)(1-z) +i\lambda[\vec\Delta \vec{k}]\bigr]
+ {q_+ \over q_+}\bigl[-m^2 +(\vec{k}_2\vec{k}_3) +i\lambda[\vec{k}_3\vec{k}_2]\bigr]
\nonumber\\
&&+ {\vec\Delta^2 -M^2 \over s} p_- 2z(1-z)q_+\Biggr\}
\nonumber\\
&& = - {1 \over \sqrt{z(1-z)}} 2z(1-z)M\label{amp7}
\end{eqnarray}

\subsection{Final trace calculation}

Once we have the expressions for vertex amplitudes,
the rest is done quickly. We first note that
each gluon vertex attached to the lower or upper line gives factor
\begin{equation}
2zq_+\cdot p_- = sz\,; \quad 2(1-z)q_+\cdot p_- = s(1-z)\label{amp8}
\end{equation}
correspondingly.

So. let's start with $T \to T$ amplitude and calculate it for Diagr.(c)
at Fig.\ref{main2}.

Equal $q\bar q$ helicities give
\begin{equation}
s(1-z)\cdot sz\cdot {-1 \over \sqrt{z(1-z)}}mV^*(\lambda)
{1 \over \sqrt{z(1-z)}}me(-\lambda)
= -s^2 m^2 e(-\lambda)V^*(\lambda)\label{amp9}
\end{equation}
Summing over $\lambda$ gives
\begin{equation}
2s^2 m^2 (\vec{e}\vec{V}^*)\,.\label{amp10}
\end{equation}
The opposite helicities yield
\begin{equation}
s^2 \left[(\vec{V}^*\vec{k})(1-2z) + i\lambda[\vec{V}^*\vec{k}]\right]
\left[(\vec{e}\vec{k}_1)(1-2z) - i\lambda[\vec{e}\vec{k}_1]\right]\label{amp11}
\end{equation}
Summing over helicities and making use of identity
\begin{equation}
[\vec a\vec b][\vec c\vec d] = (\vec a\vec c)(\vec b\vec d) - 
(\vec a\vec d)(\vec b\vec c)\label{amp12}
\end{equation}
one obtains
\begin{equation}
2s^2 \left[(\vec{V}^*\vec{k})(\vec{e}\vec{k}_1)(1-2z)^2 + 
(\vec{e}\vec{V}^*)(\vec{k}\vec k_1) - (\vec{e}\vec{k})(\vec{V}^*\vec k_1)\right]\,.
\label{amp13}
\end{equation}

Since we factored out $2s^2$ when deriving (\ref{gv4}),
we finally get
\begin{equation}
I^{(c)}_{T\to T} =
- \left[(\vec{e}\vec{V}^*)(m^2 + \vec{k}\vec k_1) + (\vec{V}^*\vec{k})(\vec{e}\vec k_1)(1-2z)^2 - (\vec{e}\vec{k})(\vec{V}^*\vec k_1)
\right]\,.\label{amp14}
\end{equation}
We included factor $(-1)$ since in this diagram we have one antiquark propagator.

An important observation here is that all other integrands, namely
$I^{(a)}\cdot (1-z)/z, I^{(c)}, I^{(d)}\cdot z/(1-z)$ give absolutely the same
result (with their own definitions of $\vec k_1$ of course). The only
thing one should not forget is that diagrams (a,d) enter with sign
'$-$' while diagrams (b,c) enter with sign '$+$':
$$
-{1-z \over z}I^{(a)} = I^{(b)} = I^{(c)} = -{z \over 1-z}I^{(d)}\,.
$$

For $L \to L$ amplitude one has immediately
\begin{equation}
I^{(c)}_{L\to L} = 
-{1 \over Q}[m^2 + \vec k_1^2 - z(1-z)Q^2]\cdot {1 \over M}2z(1-z)M^2\label{amp15}
\end{equation}
In fact, using simple relation
\begin{equation}\label{trick}
  {m^2 + \vec k_1^2 - z(1-z)Q^2 \over m^2 + \vec k_1^2 + z(1-z)Q^2} =
  1 + { - 2z(1-z)Q^2 \over m^2 + \vec k_1^2 + z(1-z)Q^2}
\end{equation}
and noting that all unity terms will eventually cancel out in
(\ref{gv17}), one can rewrite (\ref{amp15}) as
\begin{equation}
I^{(c)}_{L\to L} = -4QMz^2(1-z)^2\,.\label{amp15a}
\end{equation}

For $T \to L$ amplitude one has
\begin{equation}
I^{(c)}_{T\to L} = 2z(1-z)M(\vec{e}\vec k_1)(1-2z)\label{amp16}
\end{equation}
and for $L \to T$ amplitude one has
\begin{equation}
I^{(c)}_{L\to T} = 
{1 \over Q}[m^2 + \vec k_1^2 - z(1-z)Q^2](1-2z)(\vec{V}^*\vec{k})\,.\label{amp17}
\end{equation}
The same transformation as in $L\to L$ amplitude, leads to
\begin{equation}
I^{(c)}_{L\to T} = - 2z(1-z)Q^2(1-2z)(\vec{V}^*\vec{k})\,.\label{amp17a}
\end{equation}
 Note that in the
last three amplitudes only opposite $q \bar q$ helicities
contributed.

\newpage

\section{Averaging over $\Omega_{{\bf p}}$ for $D$ wave states}
\label{apc}

Before doing this, let's have a table of useful relations:
\begin{eqnarray}
&&\langle \vec{k}^2 \rangle = {2 \over 3}{\bf p}^2 \quad
\langle p_z^2 \rangle  =  {1 \over 3}{\bf p}^2 \nonumber\\[1mm]
&&\langle \vec{k}^2 \vec{k}^2 \rangle  =  {8 \over 15}{\bf p}^4 \quad
\langle \vec{k}^2 p_z^2 \rangle  =  {2 \over 15}{\bf p}^4 \quad
\langle p_z^2 p_z^2 \rangle  =  {3 \over 15}{\bf p}^4 \label{i2}
\end{eqnarray}
Finally, remember that $p_z^2 = {1 \over 4}(1-2z)^2M^2$.

One has to perform the following avegaring
\begin{equation}
\left\langle 4z(1-z)\cdot{1 \over \overline Q^4}\cdot
\left(\vec{k}^2 - {4m\over M}p_z^2\right)\cdot \left(1 - {4 \vec{k}^2\over \overline Q^2}
\right)\right\rangle
\label{i15ap}
\end{equation}
Note that all factors should be carefully examined; all four do contribute
to the final answer.
Decomposing $\overline Q^2$ as
\begin{equation}
\overline Q^2 = m^2 + z(1-z)Q^2 = m^2 + {1 \over 4}Q^2 - {1 \over 4} (1-2z)^2Q^2
\equiv \overline Q^2_0 - {p_z^2 \over M^2}Q^2\,,\label{i16ap}
\end{equation}
one gets
\begin{equation}
{1 \over \overline Q^4} = {1 \over \overline Q_0^4}
\left( 1 + 2{p_z^2 \over M^2}{Q^2 \over \overline Q_0^2}\right)\,.\label{i17ap}
\end{equation}
So, omitting $\overline Q_0^{-4}$, one has
\begin{equation}
\left\langle \left(1 -{4p_z^2 \over M^2}\right)
\cdot\left(1 + 2{p_z^2 \over M^2}{Q^2 \over \overline Q_0^2}\right)
\cdot\left(\vec{k}^2 - {4m\over M}p_z^2\right)\cdot \left(1 - {4 \vec{k}^2\over \overline Q_0^2}
\right)\right\rangle
\label{i18ap}
\end{equation}
With the aid of (\ref{i1}), one obtains
\begin{eqnarray}
&&\left\langle \vec{k}^2 - {4m\over M}p_z^2 \right\rangle
- {4 \over M^2}\left\langle \vec{k}^2 p_z^2 - 2 p_z^2 p_z^2 \right\rangle
+ 2{Q^2 \over M^2 \overline Q^2_0}\left\langle \vec{k}^2 p_z^2 - 2 p_z^2 p_z^2 \right\rangle
- {4 \over \overline Q_0^2} \left\langle \vec{k}^2 \vec{k}^2 - 2 p_z^2 \vec{k}^2\right\rangle
\nonumber\\[1mm]
&=&\left({2 \over 3}{\bf p}^2  - {4m \over 3M}{\bf p}^2 \right)
- {4 \over M^2} {\bf p}^4 \left( {2 \over 15} - {6 \over 15}\right)
+  2{Q^2 \over M^2 \overline Q^2_0}{\bf p}^4\left( {2 \over 15} - {6 \over 15}\right)
- {4 \over \overline Q_0^2} {\bf p}^4\left( {8 \over 15} - {4 \over 15}\right)
\nonumber\\[1mm]
&=&{2 \over 3}{\bf p}^2{4 {\bf p}^2 \over M+2m} + {16{\bf p}^4 \over 15 M^2}
- { 8Q^2{\bf p}^4 \over 15 M^2 \overline Q^2_0}
-{16{\bf p}^4 \over 15 \overline Q^2_0}\label{i19ap}\\[1mm]
&=&{4 {\bf p}^4 \over 3M^2}
\left( 1 + {4 \over 5} - {8 \over 5}{Q^2 \over Q^2 + M^2}
- {16 \over 5}{M^2 \over Q^2 + M^2} \right)\nonumber\\[1mm]
&=&{4 {\bf p}^4 \over 15M^2}\left(1 - 8{M^2 \over Q^2 + M^2}\right)\nonumber
\end{eqnarray}

For the helicity conserving $T \to T$ amplitude one has to repeat the same
averaging procedure. The expression to be calculated first is
\begin{eqnarray}
&&\Biggl\langle \left(1 + 2{p_z^2 \over M^2}{Q^2 \over \overline Q_0^2}\right)
\cdot
\Biggr[ 2{\bf p}^2\left( m^2 + 2\vec{k}^2 - 4\vec{k}^2 {m^2 \over \overline Q^2}\right)
-m(M+m)\vec{k}^2 \left(1 - {4 \vec{k}^2 \over \overline Q^2}\right)
-2\vec{k}^2 \left(\vec{k}^2 - {4m \over M}p_z^2\right)\Biggr]\Biggr\rangle
\nonumber\\
&& = 2 m^2 {\bf p}^2 - {2 \over 3}m(M+m){\bf p}^2
+ 2{p_z^2 \over M^2}{Q^2 \over \overline Q_0^2}
\left({1\over 2}M^2{1\over 3}- 3 M^2{2 \over 15} \right)
+ {8 \over 3}{\bf p}^4 -   {8 \over 15} {\bf p}^4
+ {m^2 \over \overline Q^2_0}{\bf p}^4{16 \over 15}\nonumber\\
&&= - {2\over 3}{\bf p}^4 - {8\over 15}{\bf p}^4 + {8\over 3}{\bf p}^4
+ {16 \over 15} {M^2 \over M^2 + Q^2}{\bf p}^4
+ {8 \over 15} {Q^2 \over M^2 + Q^2}{\bf p}^4
\nonumber\\
&&= 2{\bf p}^4 \left(1 + {4 \over 15} {M^2 \over M^2 + Q^2}\right)\,.
\label{i23ap}
\end{eqnarray}

\newpage

{\Large
\begin{center}
{\bf Acknowledgements}
\vspace{1cm}
\end{center}
}

The author wishes to thank Prof. J.Speth for the
opportunity to be accepted as a collaborator at
Theoretical particle physics group at Forschungszentrum Juelich and
Prof. I.F.Ginzburg, who made this visit possible.
I am truly grateful to Prof. N.N.Nikolaev
for his wise and patient guidance through the entire work.
I wish also thank the other colleagues,
in particular W.Sch\"{a}fer, who spent their time
answering my questions or asking theirs.
Besides, I am also grateful to Profs. V.G.Serbo
and V.S.Fadin at Novosibirsk University
for their wonderful lecture courses
"Two photon interactions" and "Introduction to QCD",
without which my comprehension of these topics
would be much more staggering and painful.
Finally, I thank all those people who in various ways
made this work possible.

\newpage

{\Large
\begin{center}
{\bf Curriculum Vitae}
\vspace{1cm}
\end{center}
}
\begin{enumerate}
\item Name - {\bf \large Igor Ivanov}
\item Date of birth - 26 October 1976
\item Place of birth -  Petropavlovsk-Kamchatsky,  Russia
\item Educational and scientific background:\hspace{-1in}
\begin{itemize}
\item[---]  graduated from Physics and Mathematics School at Novosibirsk University
           in June 1993;
\item[---]  received Physics  Bachelor Diploma of Novosibirsk University in July 1997.
         My Bachelor Diploma Research  was based on ref.[1].
\item[---] 1998--1999: graduate student (Studentenhilfskraft) at
         Instit\"{u}t f\"{u}r Kernphysik, Forschungszentrum
         J\"{u}lich, Germany
\item[---] received MSc Diploma in Physics of Novosibirsk University in June 1999.
\item[---] currently PhD student of Novosibirsk University and Doktorand
at  Instit\"{u}t f\"{u}r Kernphysik, \\Forschungszentrum J\"{u}lich 
\end{itemize}
\item Teaching experience - I have been teaching physics seminars and the elective
        course 'Olympiad problems in physics' at the Physics and Mathematics School
        (Novosibirsk) for three years.
\item Publications:
\begin{description}
\item[{[1]}]    I.F.Ginzburg, I.P.Ivanov  {\it  'Higgs boson search at photon colliders
                   for $M_H=140-190$ GeV'}, {\it Phys.Lett.} {\bf B408} (1997) 325;
                   {\bf hep-ph/9704220}.
\item[{[2]}]    I.F Ginzburg, I.P.Ivanov, A. Schiller   
{\em 'Search for Next Generations of Quarks
 and Leptons at the Tevatron and LHC'}, 
to appear in {\it Phys. Rev. D}; {\bf hep-ph/9802364}.
\item[{[3]}]    A.T.Banin, I.F.Ginzburg, I.P.Ivanov  {\it 'Anomalous interactions in Higgs boson
                   production at $\gamma\gamma$ and $e\gamma$ colliders'}, {\it Phys. Rev.}
                   {\bf D 59} (1999) 115001; {\bf hep-ph/9806515}.
\item[{[4]}] I.P.Ivanov, N.N.Nikolaev, {\it Pis'ma ZhETF (JETP Lett.)}
        {\bf 69} (1999) 268; {\bf hep-ph/9901267}.
\item[{[5]}] I.P.Ivanov, N.N.Nikolaev, A.~Pronyaev, W.~Schaefer,
        {\it Phys. Lett.}, {\bf B 457} (1999), 218-226 ;{\bf hep-ph/9903228}. 

\end{description}

\item Fields of interest:  quantum field theory, high energy physics,
Standard model and beyond, Higgs boson physics,
high energy photon interactions, QCD, diffractive deep-inelastic processes,
spin structure functions.
\end{enumerate}

\end{document}